\documentclass[preprintnumbers,superscriptaddress,nofootinbib,aps,prd,twocolumn,floatfix]{revtex4}
\pdfoutput=1

\usepackage{enumerate}
\usepackage{graphicx}
\usepackage{amsmath}
\usepackage{amssymb}
\usepackage{color}
\usepackage[yyyymmdd,hhmmss]{datetime}
\usepackage[normalem]{ulem}
\usepackage[multiple]{footmisc}
\usepackage{booktabs}
\usepackage{multirow}
\usepackage{pbox}
\usepackage{tabularx}
\usepackage{slashed}
\usepackage{snapshot}

\newcommand{\newc}{\newcommand}
\def\beq{\begin{equation}}
\def\eeq{\end{equation}}
\def\beqn{\begin{eqnarray}}
\def\eeqn{\end{eqnarray}}
\newc{\gluino}{\tilde g}
\newc{\Conep}{\tilde \chi_1^+}
\newc{\Conem}{\tilde \chi_1^-}
\newc{\Cone}{\tilde \chi_1^\pm}
\newc{\Ntwo}{\tilde \chi_2^0}
\newc{\None}{\tilde \chi_1^0}
\newc{\stopp}{\tilde t}
\newc{\stopa}{\tilde t^*}
\newc{\stopone}{\tilde t_1}
\newc{\stopones}{\tilde t_1^*}

\newcommand{\gev}{{\ensuremath\rm GeV}}

\newcommand\be{\begin{eqnarray}}
\newcommand\ee{\end{eqnarray}}
\newcommand\ben{\begin{eqnarray*}}
\newcommand\een{\end{eqnarray*}}
\newc{\red}{\textcolor{red}}
\newcommand{\amc}{{\sc MadGraph5\textunderscore}a{\sc MC@NLO}}

\begin{document}
\preprint{IPMU17-0146, PITT-PACC-1713}

\title{Mono-top Signature from Fermionic Top-partner}

\author{Dorival Gon\c{c}alves}
\affiliation{PITT-PACC,  Department  of  Physics  and Astronomy,  University  of  Pittsburgh,  USA}
\author{Kyoungchul Kong}
\affiliation{Department of Physics and Astronomy, University of Kansas, Lawrence, KS 66045, USA}
\author{Kazuki Sakurai}
\affiliation{Institute of Theoretical Physics, Faculty of Physics, University of Warsaw,  Poland}
\author{Michihisa Takeuchi}
\affiliation{Kavli IPMU (WPI), UTIAS, University of Tokyo, Kashiwa, 277-8583, Japan}

\begin{abstract}
  \noindent
We investigate mono-top signatures arising from phenomenological models of fermionic top-partners, which are degenerate
in mass and decay into a bosonic dark matter candidate, either spin-0 or spin-1. Such a model provides a mono-top signature 
as a smoking-gun, while conventional searches with $t\bar t$ + missing transverse momentum are limited. Two such 
scenarios: 
i) a phenomenological 3rd generation extra dimensional model with excited top and electroweak sectors,
and ii) a model where only a top-partner and a dark matter particle are added to the SM, 
are studied in the degenerate mass regime.
We find that in the case of extra dimension a number of different processes give rise to effectively the same mono-top
final state, and a great gain can be obtained in the sensitivity for this channel. We show that the mono-top search can 
explore top-partner masses up to
 630~GeV and 300~GeV for the 3rd generation extra dimensional model and the minimal fermionic top-partner model, 
 respectively, at the high luminosity LHC.
\end{abstract}
\preprint{}

\maketitle


\newpage
\section{Introduction}\label{sec:intro}


Mono-top searches have been proposed in a context of supersymmetry (SUSY)~\cite{Goncalves:2016tft, Goncalves:2016nil, 
Fuks:2014lva},\footnote{ The mono-top signatures have also been studied in context of flavor models~\cite{Aad:2014wza, 
Khachatryan:2014uma, Davoudiasl:2011fj, Kamenik:2011nb, Andrea:2011ws, Alvarez:2013jqa, Agram:2013wda, Boucheneb:2014wza}.} 
especially when superpartner (stop, $\tilde t$) of the Standard Model (SM) top quark ($t$) is degenerate with (higgsino-like) neutralinos ($\tilde h^0$).
In such a scenario, $\tilde t$ effectively behaves as an invisible particle since its decay products are too soft to be detected,
and the sensitivity is therefore deteriorated for the standard $pp \to {\tilde t} \tilde t^\ast$ channel~\cite{Aaboud:2016tnv, Khachatryan:2016pxa}. 
It has been shown that in this regime the $pp \to {\tilde t} t \tilde h^0$ channel can have a measurable production rate due to the large top Yukawa coupling, 
leading to a characteristic mono-top~+~$\slashed{E}_T$ final state, where both $\tilde t$ and $\tilde h^0$ are effectively invisible. 
Differently than the usual mono-jet signatures exploiting hard QCD initial state radiation and therefore providing very little information on the produced particles, the mono-top signature allows a direct probe of the stop and neutralino sectors \cite{Goncalves:2016tft}.
This means that the ${\tilde t} t \tilde h^0$ channel is complementary to the mono-jet channel or even essential 
in exploring the SUSY parameter space.
While such a supersymmetric signature is very well motivated theoretically in terms of naturalness
~\cite{Kitano:2006gv,
Papucci:2011wy,
Hall:2011aa,
Desai:2011th,
Ishiwata:2011ab,
Sakurai:2011pt,
Kim:2009nq,
Wymant:2012zp,
Baer:2012up,
Randall:2012dm,
Cao:2012rz,
Asano:2012sv,
Baer:2012uy,
Evans:2013jna,
Hardy:2013ywa,
Kribs:2013lua,
Bhattacherjee:2013gr,
Rolbiecki:2013fia}, 
it is desirable to phenomenologically expand the scope of current mono-top studies including different spin-scenarios. As supersymmetry provides a spin-0 top 
partner together with a spin-$1/2$ dark matter (DM) candidate, we would like to extend the search to the case with a spin-$1/2$ top-partner, which 
decays into either spin-0 or spin-1 DM candidate. As in the supersymmetric case, we will assume the degeneracy between the fermionic top-partner and bosonic 
DM candidates.

For spin-0 top-partner case, `natural SUSY' provides a well-motivated example for the degenerate spectrum. Similarly, for spin-1/2 top partner case, such a compressed spectrum is naturally realized in models with extra dimensions. 
A good benchmark model for the purpose of our analysis is Universal Extra Dimensions (UED), where all SM particles propagate in the bulk of flat extra dimensions, and the mass spectrum of Kaluza-Klein (KK) particles is 
degenerate~\cite{Appelquist:2000nn}. This degeneracy is broken due to electroweak symmetry breaking, bulk and boundary term corrections from the renormalization group running between the cut-off scale and the electroweak scale \cite{Georgi:2000wb,Cheng:2002ab,Cheng:2002iz,Flacke:2013pla,Flacke:2017xsv}. Nevertheless, overall mass spectrum is much 
narrower than that for conventional supersymmetry. This observation strongly encourages revisiting mono-top production in the context of extra 
dimensions. 

UED with a particular mass spectrum derived in \cite{Cheng:2002ab,Cheng:2002iz} is called as `Minimal Universal Extra Dimensions' (MUED) and has been extensively studied in the literature \cite{Choudhury:2016tff,Deutschmann:2017bth,Beuria:2017jez,Georgi:2000wb,Cheng:2002ab,Cheng:2002iz,Flacke:2013pla,Flacke:2017xsv,Belanger:2010yx,Servant:2002hb,Kong:2005hn,Cheng:2002ej,Arrenberg:2008wy,Datta:2005zs}.
Recently, it has been revisited with the 8 TeV and 13 TeV LHC data with the conventional cascade decays \cite{Deutschmann:2017bth,Beuria:2017jez}, and the estimated lower bound on the inverse radius is found to be around $R^{-1}\gtrsim~1.4$~TeV with some variation in the cut-off scale ($\Lambda$). These searches with jets, leptons and missing transverse momentum are promising in general, but their sensitivity gets poor for smaller mass splitting. On the other hand, the mono-jet channel becomes more sensitive for compressed spectra, which is expected for a low value of $\Lambda R$. This point is examined in Ref.~\cite{Choudhury:2016tff}, and they find that mono-jet searches result in the current bound $R^{-1} \gtrsim 1.1$~TeV ($\sim$~1.2 TeV and $\sim$ 1.3 TeV for the masses of KK quark and KK gluon, respectively) for $\Lambda R \lesssim$ 5 with $3.2$ fb$^{-1}$ of data at the 13 TeV LHC, 
which is comparable to the mono-jet exclusion limits on the masses of squarks ($\gtrsim 0.8$~TeV) and gluino ($\gtrsim 1$~TeV) in supersymmetry~\cite{ATLAS-CONF-2017-022,ATLAS-CONF-2017-060}.
Since the mono-top signals arise from 3rd generation of KK quarks, we will study the mono-jet channel with the corresponding particle content in our study. This should be compared to Ref.~\cite{Choudhury:2016tff}, where entire KK spectrum is included in the analysis. 

Although collider phenomenology of extra dimensional models has been examined extensively for many years, 
its mono-top signature has not been pursued yet. 
Therefore, our study is worthwhile and will provide useful information concerning SUSY and UED searches. Moreover, in a particular case where only KK tops and KK dark matter candidates are considered without additional KK particles, our analysis is more generally applicable beyond extra dimensional models and our results would be still valid in a generic model with fermionic top partners and a dark matter candidate. 
Hence, although, in this paper, we refer to the fermionic top-partner as KK top (denoted by $t^{(1)}$ or $T^{(1)}$, depending on their $SU(2)_W$ charge), 
and the bosonic DM candidate as KK photon ($\gamma_\mu^{(1)}$) for spin-1 or KK Higgs ($h^{(1)}$ or $\chi^{(1)}$, depending on their CP property) for spin-0, our results can be more general.
We consider KK number conserving interactions in our study and therefore all interactions are fixed by the SM ones.
Masses of the new particles are treated as free parameters, as in non-minimal UED \cite{Flacke:2013pla}, which we fix following the previous SUSY studies for comparison~\cite{Goncalves:2016tft,Goncalves:2016nil}. We also take 
SUSY decay chains used in the previous study to guarantee an appropriate comparison against the existing results. All SUSY particles will be replaced with the corresponding `KK' partners with different spins.

This paper is organized as follows. In Sec.~\ref{sec:model}, we define two benchmark scenarios which are addressed in our study and describe the top-partner interactions that are relevant to the mono-top signature. 
In Sec.~\ref{sec:analysis}, we present the result of our numerical study based on
Monte Carlo simulations and derive the corresponding LHC bounds. 
Finally, a summary of our main findings is given in Sec.~\ref{sec:conclusion}.

\section{Relevant interactions for fermionic top-partner}
\label{sec:model}

We consider two different scenarios in our mono-top study: 
\begin{enumerate}[(i)]
\item {\it a phenomenological 3rd generation extra dimensional model, which consists of 
the top-partner sector 
($SU(2)_W$ singlet KK top $t^{(1)}$, and third generation $SU(2)_W$ doublet, $(T^{(1)}, B^{(1)})$)  
as well as the full KK Higgs and KK electroweak gauge boson spectrum. 
Such a scenario may be realized in non-minimal UED models \cite{Datta:2016flx,Biswas:2017vhc,Flacke:2008ne,Ishigure:2016kxp,Flacke:2013pla,Huang:2012kz}. }

\item {\it a minimal scenario with one fermionic top-partner $\,t^{(1)}$ and bosonic DM candidate 
$h^{(1)}$ (spin-0) or $\gamma_\mu^{(1)}$ (spin-1), as to stop plus neutralino corresponding to the simplified model in SUSY.}
\end{enumerate}
%
Following particle content and interactions as in UED, 
we assume the lightest KK particle (LKP) to be electrically neutral and colorless,
so as to be the dark matter candidate. 
As long as the LKP is stable and invisible within the detectors, 
further specification of the LKP is not important 
since decays of KK particles are not visible due to the mass-degeneracy among them.

The KK photon $\gamma_\mu^{(1)}$ is essentially the KK hyper-charge gauge boson,
 $\gamma_\mu^{(1)} \approx B_\mu^{(1)}$, since the Weinberg angle at KK level is small, $\theta_W^{(n)} \ll 1$.
Similarly the KK $Z$ consists of mostly neutral component of $SU(2)_W$ KK gauge boson, $Z_\mu^{(1)} \approx W_\mu^{(1)3}$. 
This is analogous to the case of pure bino $\tilde b$ and zino $\tilde z$ in SUSY. 
$W_\mu^{(1)\pm}$ and $H^{(1)\pm}$ are the charged KK $W$ and KK Higgs bosons. We denote CP even and CP odd Higgs bosons 
as $h^{(1)}$ and $\chi^{(1)}$, respectively. 
The SM top quark and KK top quarks form the following Dirac fermions:
\begin{eqnarray}
t &=& \left ( \begin{array}{c}
T_L \\
t_R
\end{array} \right ) , ~~~~{\rm SM~ top~ quark } \, , \notag \\
T^{(1)}  &=& \left ( \begin{array}{c}
T_L^{(1)} \\
T_R^{(1)}
\end{array} \right ) , ~~SU(2)_W~ {\rm doublet~KK~top} \, ,   \\
t^{(1)}  &=& \left ( \begin{array}{c}
t_L^{(1)} \\
t_R^{(1)} 
\end{array} \right ) , ~~~ \,SU(2)_W~ {\rm singlet~KK~top} \, . \notag
%
%
\end{eqnarray}
%
Often $t^{(1)}$ ($T^{(1)}$) is called the right-handed (left-handed) KK top. However,  it is really a vector-like quark. 
The handedness refers to the chirality of the SM fermion of its origin, {\it i.e.,} $t^{(1)}$ is KK partner of the right-handed SM top $t_R$ and $T^{(1)}$ is the KK partner of the left-handed SM top $T_L$. 
%
The relevant interactions involving the SM top quark and the KK electroweak gauge bosons are
\begin{eqnarray}
g_1 \, \frac{Y_R}{2}    \, \bar{t}^{(1)} \, \gamma^\mu P_R \, t \,  \gamma_\mu^{(1)} \,  + h.c.  \, ,\\
g_1 \, \frac{Y_L}{2}     \, \bar{T}^{(1)} \, \gamma^\mu P_L \, t \, \gamma_\mu^{(1)} \,  + h.c.  \, ,\\
g_2 \, \frac{1}{2}          \, \bar{T}^{(1)} \, \gamma^\mu P_L \, t \, Z_\mu^{(1)} \,  + h.c.  \, ,\\
g_2 \, \frac{1}{\sqrt{2}} \, \bar{B}^{(1)} \, \gamma^\mu P_L \, t \, W_\mu^{(1)-} \,  + h.c.  \, , 
\end{eqnarray}
%
where $Y_{L/R}$ is the corresponding hyper-charge, and $g_1$ and $g_2$ are the gauge coupling strengths of $U(1)_Y$ and $SU(2)_W$ interactions,
respectively ($Y_R = 4/3$ and ${Y_L =1/3}$).

The $SU(2)_W$ doublet fields are defined as 
\begin{widetext}
\begin{eqnarray}
q_L &=& \left (
\begin{array}{c}
T_L \\ B_L
\end{array}
    \right ), \hspace*{0.1cm}
q^{(1)} = \left (
\begin{array}{c}
T^{(1) }\\ B^{(1)}
\end{array}
    \right )    , \hspace*{0.1cm}
H = \left (
\begin{array}{c}
H^+ \\ \frac{1}{\sqrt{2}} \left ( v + h + i \chi \right )
\end{array}
    \right )    , \hspace*{0.1cm}
H^{(1)} = \left (
\begin{array}{c}
H^{(1)+} \\ \frac{1}{\sqrt{2}} \left ( h^{(1)} + i \chi^{(1)} \right )
\end{array}
    \right )    \, ,
    \label{eq:lag2}
\end{eqnarray}
\end{widetext}
where $f_{R/L}=P_{R/L} f$ for a fermion $f$.

The interactions involving the SM top quarks and KK Higgs are given by 
\begin{widetext}
\begin{eqnarray}
&&\hspace*{-0.5cm}{\cal L} \ni \lambda_t \left [ 
	   \bar q_L t_R i \sigma_2 H^\ast 
	+ \bar q_L t_R^{(1)} i \sigma_2 H^{(1)\ast}
	+ \bar q_L^{(1)} t_R i \sigma_2 H^{(1)\ast}
\right ] +
\lambda_b \left [ 
	   \bar q_L b_R  H  
	+ \bar q_L b_R^{(1)}  H^{(1)}
	+ \bar q_L^{(1)} b_R H^{(1)}
\right ] + h.c.  \notag
 \\
&&\hspace*{-0.5cm}= \lambda_t  \left [  
	   \bar T_L t_R \frac{1}{\sqrt{2}} (v + h) 
	+ \bar T_L t_R^{(1)} \frac{1}{\sqrt{2}} ( h^{(1)} - i \chi^{(1)}) 
	 - \bar B_L t_R^{(1)} H^{(1)-}  
	 - \bar B_L^{(1)} t_R H^{(1)-}
	+ \bar T_L^{(1)} t_R \frac{1}{\sqrt{2}} ( h^{(1)} - i \chi^{(1)}) 
\right ] + 
  \\
&& \hspace*{-0.5cm} \lambda_b \left [  
	   \bar B_L b_R \frac{1}{\sqrt{2}} (v + h) 
	+ \bar B_L b_R^{(1)} \frac{1}{\sqrt{2}} ( h^{(1)} + i \chi^{(1)}) 
	 + \bar T_L b_R^{(1)} H^{(1)+}  
	 + \bar T_L^{(1)} b_R H^{(1)+}
	+ \bar B_L^{(1)} b_R \frac{1}{\sqrt{2}} ( h^{(1)} + i \chi^{(1)}) 
\right ] + h.c.  \, , \notag
\label{eq:lag1}
\end{eqnarray}
\end{widetext}
where $m_{t(b)} = \frac{\lambda_{t(b)} v}{\sqrt{2}}$,  and $v = \frac{2 m_W}{g_2}\approx 246$ GeV are the top (bottom) quark mass 
and the Higgs vacuum expectation value, respectively.

\section{Analysis}
\label{sec:analysis}

\begin{figure}[t!]
\includegraphics[width=0.50\columnwidth]{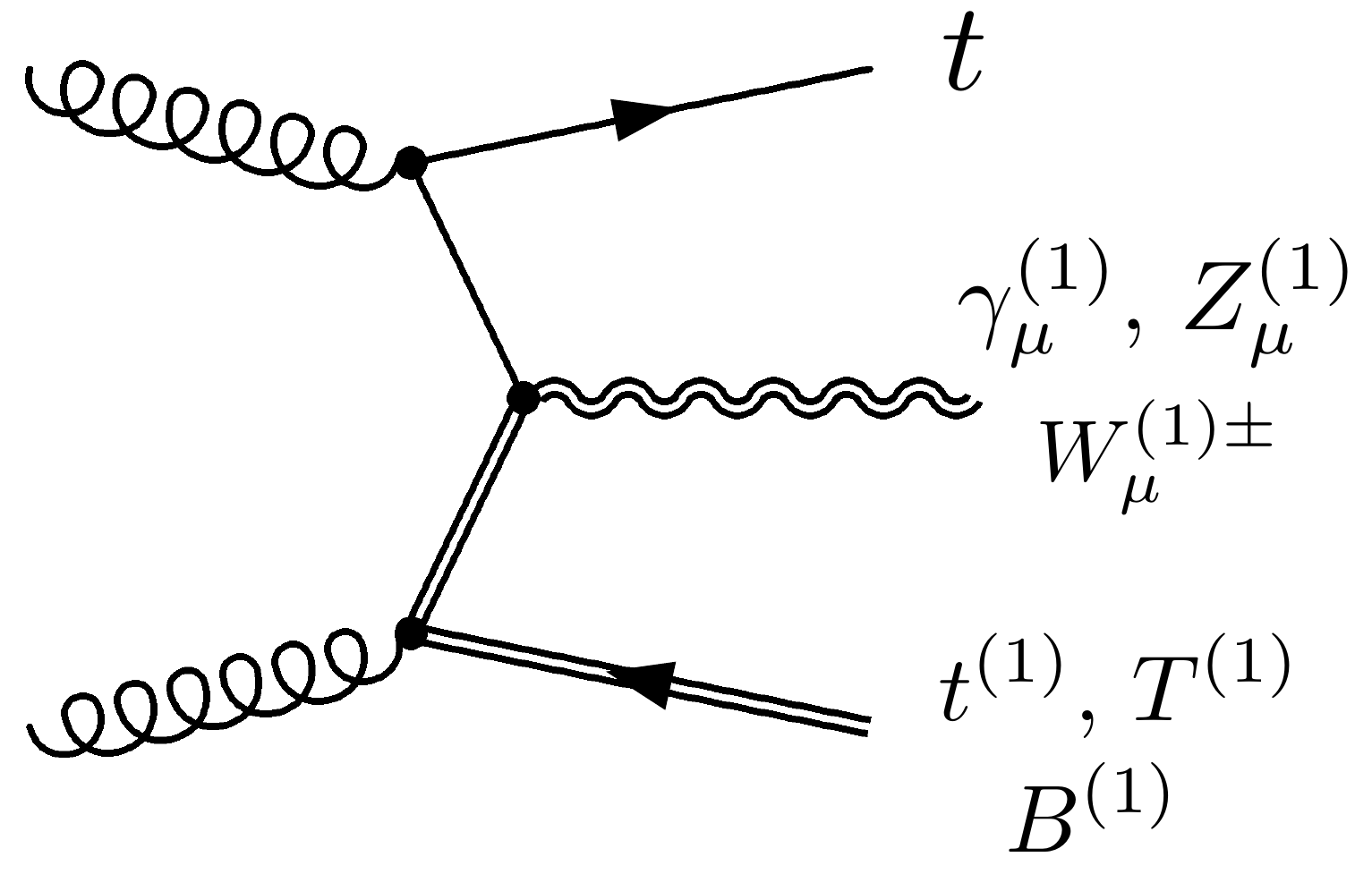}\hspace{-0.15cm}
\includegraphics[width=0.50\columnwidth]{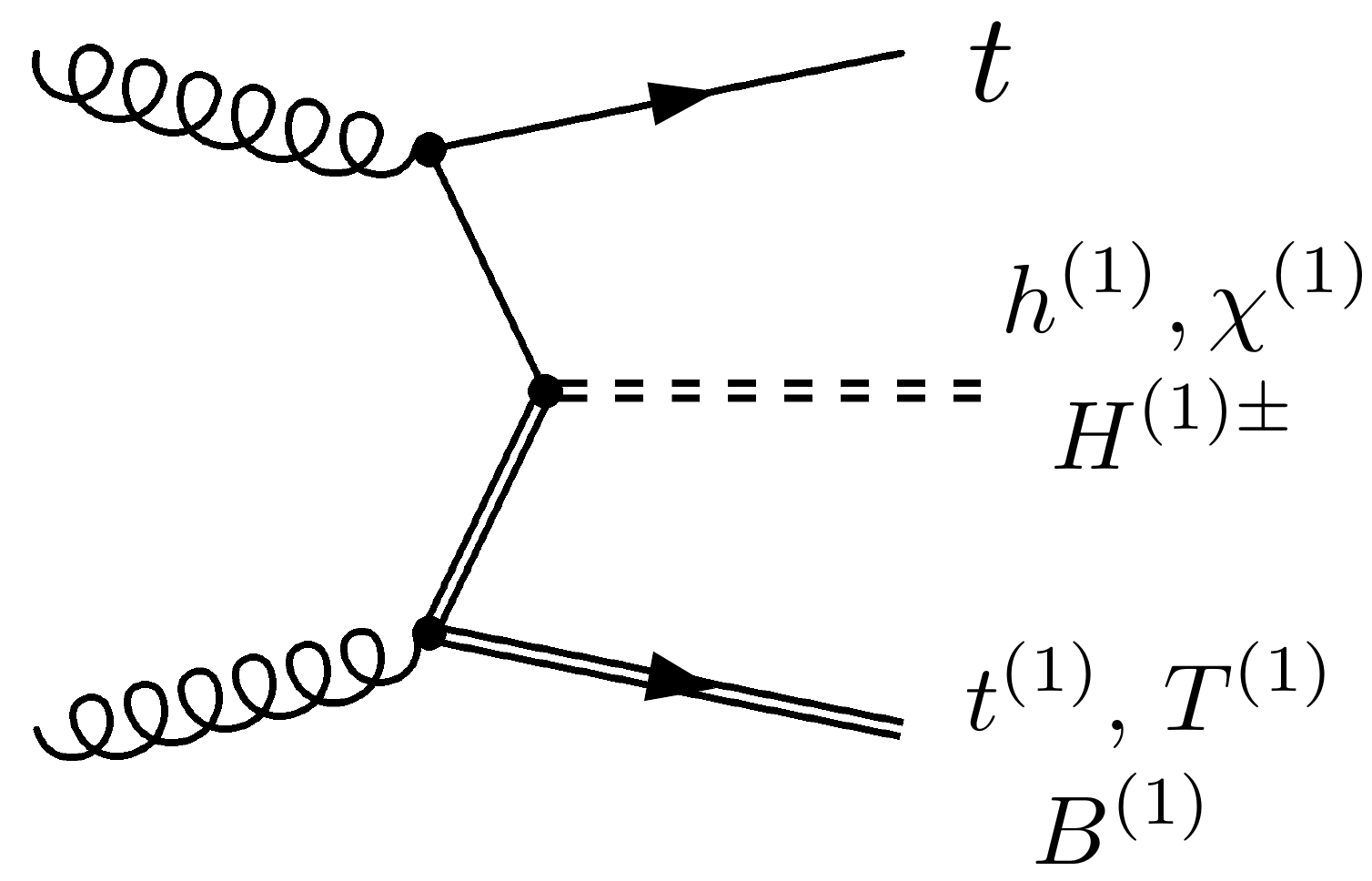}
\caption{Representative set of Feynman diagrams resulting in the mono-top signature
in the 3rd generation extra dimensional model. 
 \label{fig:diagrams}}
\end{figure}

We consider the mono-top signature arising from $(i)$ the 3rd generation extra dimensional model ${pp \to t\, {\rm KK}_f^{(1)} {\rm KK}_b^{(1)}}$ and $(ii)$ the minimal top-partner scenario  ${pp \to t\, t^{(1)} h^{(1)}}$, where ${\rm KK}_f^{(1)}$ represents any 3rd generation KK quark $t^{(1)}$, $T^{(1)}$ or $B^{(1)}$, and ${\rm KK}_b^{(1)}$ any KK boson that couples to the top quark, as illustrated  in Fig.~\ref{fig:diagrams}.\footnote{The KK gluon is not included here, since it is often the heaviest particle in UED models.}
 In scenario $(i)$, as illustrated in the figure, many different processes effectively contribute to the same mono-top + $\slashed{E}_T$ 
final state, as all KK particles are quasi-mass-degenerate, being essentially invisible with soft and undetected decay products. 
Following Refs.~\cite{Goncalves:2016tft,Goncalves:2016nil}, we set in our analysis 
$m_{\gamma^{(1)}}=m_{Z^{(1)}}=m_{\chi^{(1)}} = m_{h^{(1)}}$,
$m_{H^{(1)\pm}}=m_{W^{(1)\pm}}=m_{\gamma^{(1)}} + 1~\gev$,
$m_{t^{(1)}}=m_{T^{(1)}}=m_{B^{(1)}}=m_{\gamma^{(1)}} + 8~\gev$,
and assume no particle 
has a detector scale lifetime
(except for the stable lightest KK particles).  

\begin{figure}[t!]
\includegraphics[width=0.65\columnwidth]{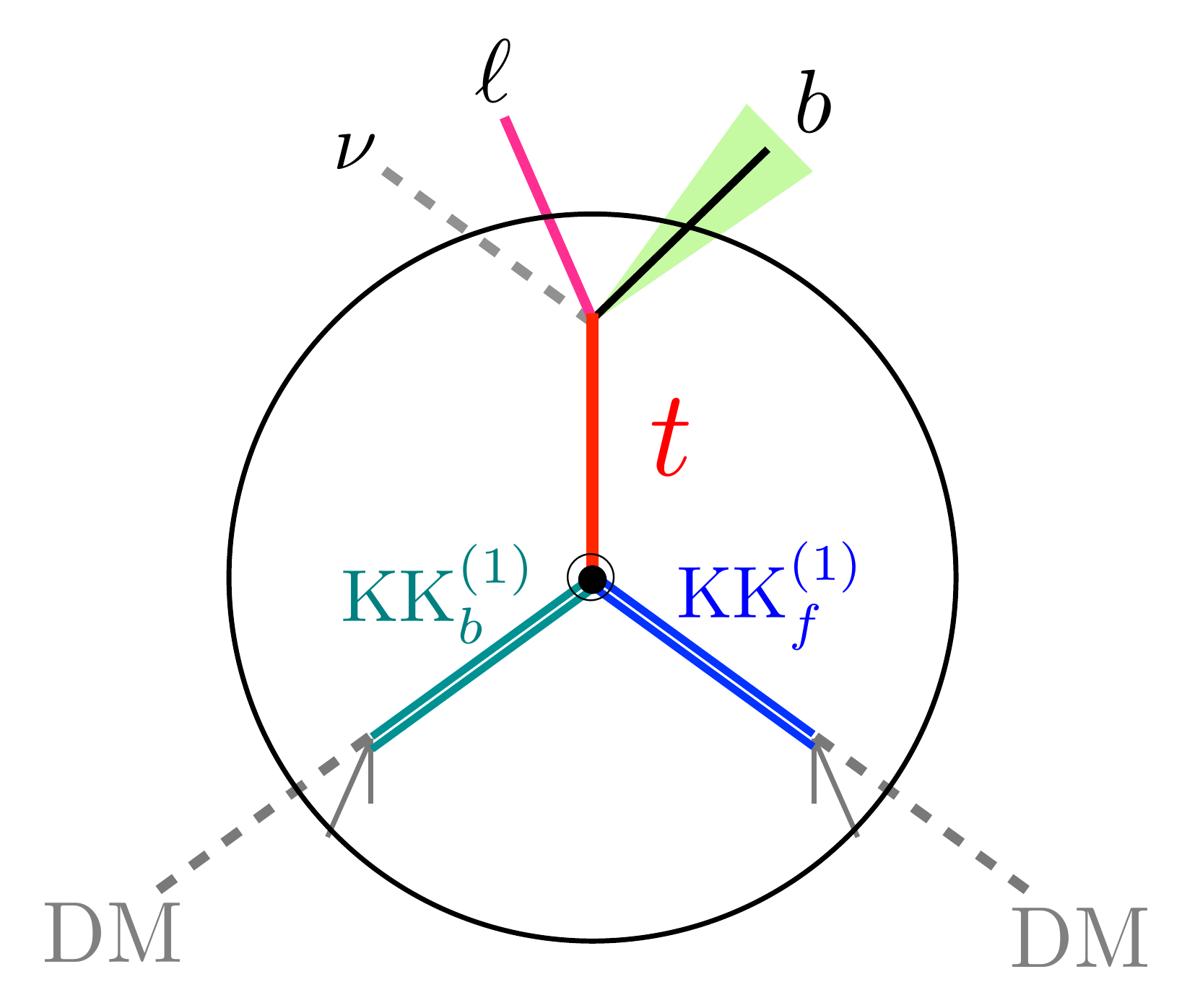}
\caption{Schematic mono-top event display. The grey dashed lines represent invisible particles,
whereas the thin grey lines depict soft particles that do not pass the minimum selection criteria.  \label{fig:event}}
\end{figure}

In this analysis, we concentrate on the leptonic mono-top signature, see Fig.~\ref{fig:event} for a schematic mono-top event display. This channel is characterized by the presence  of an isolated lepton $\ell=e,\mu$, one $b$-tagged jet and missing energy $\slashed{E}_T$. 
The dominant backgrounds for this signature are $\bar{t}t$+jets, $tW$, $tZ$ and $W\bar{b}b$ production processes.

In our analysis, we generate the $\bar{t}t$+jets sample with \textsc{ALPGEN+Pythia6}~\cite{Mangano:2002ea} 
merged up to one jet, with the \textsc{MLM} multi-jet merging algorithm. The signal and additional  
background samples are generated with \textsc{\amc+Pythia8}~\cite{mg5,pythia}, accounting for 
hadronization and underlying event effects.  Detector effects  are simulated with the
\textsc{Delphes3} package~\cite{delphes}. Higher order corrections are accounted for by normalizing the total $\bar{t}t$ 
rate to the NNLO+NNLL cross-section (831~pb~\cite{tt_NNLO}), and the $tW$ and $tZ$ to their NLO predictions~(71~pb~\cite{LHCTopWG})
and~(0.88~pb~\cite{tz}), respectively.\footnote{The literature does not provide higher order corrections to the signal under consideration. 
We indicate the importance of its determination for future studies.}
We use the K-factor of 1.5 for the signal processes.

\begin{figure}[b!]
\includegraphics[width=0.92\columnwidth]{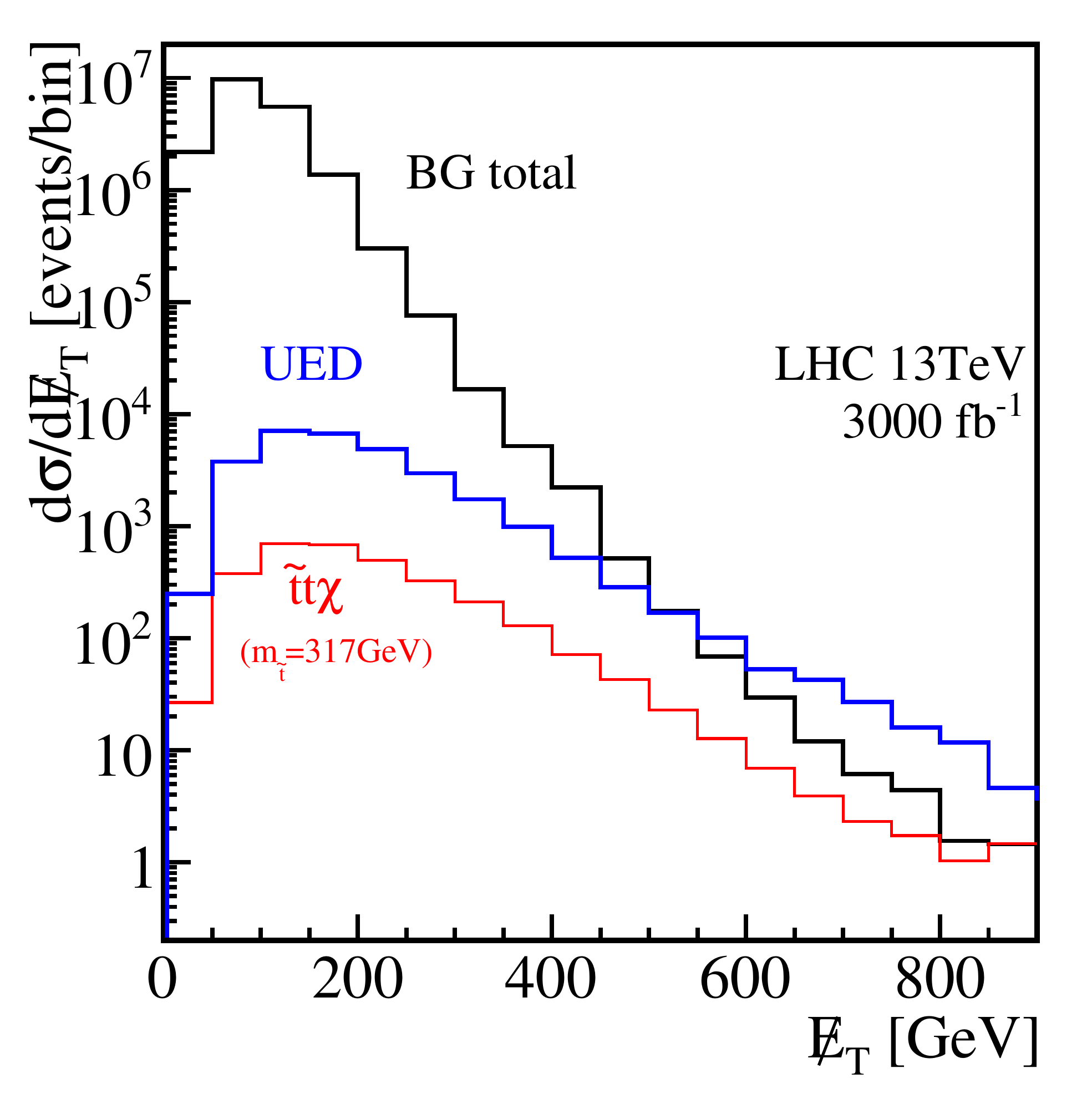}
\caption{The missing energy $\slashed{E}_T$ distributions for the SM background (black) and the mono-top 
signal in the 3rd generation extra dimensional model (case $(i)$, blue). 
The distribution for the SUSY case with the $\tilde{t} t \tilde \chi_1^{0}$ channel is also shown in red. 
For illustration purposes, we show the  SUSY and UED distributions with the same top-partner masses.
We assume the 13~TeV LHC with 3~ab$^{-1}$ of integrated luminosity.  \label{fig:etmiss}}
\end{figure}

\begin{table*}[t!]
\centering
\resizebox{13cm}{!}{
\begin{tabular}{c||r||r|r|r||r|r|r}
  Process    &  $\sigma$  &       Baseline  &  $m_{b\ell} < 150$  &  $m_T > 100$  
  &  $\slashed{E}_T>550$ & $\slashed{E}_T>600$ & $\slashed{E}_T>650$ \\
\hline
$t \bar t$   &   831\,pb  &  $206 \cdot 10^6$  &  $165. \cdot 10^6$   &  $17.7 \cdot 10^6$  
&  55.2 & 25.0 & 11.2 \\ 
$t W$        &    71\,pb  &  $26.2 \cdot 10^6$  &  $20.7 \cdot 10^6$  &  $1.68 \cdot 10^6$  
&  55.5  & 24.3 & 10.4 \\
$t Z$        &  0.88\,pb  &  $22.8 \cdot 10^3$  &  $21.6 \cdot 10^3$  &  $7.3 \cdot 10^3$   
&  8.0   & 4.9 & 3.5 \\ 
$Wb \bar b$ &    7.65\,pb  &  $1.82 \cdot 10^6$     &  $1.51 \cdot 10^6$     &  $42.3 \cdot 10^3$    
&  1.4 & 0.7 &  0.3 \\ 
\hline
BG total     &   903\,pb  &  $226 \cdot 10^6$   &  $41.1 \cdot 10^6$  &  $19.4 \cdot 10^6$  
&  120.1 & 54.9 & 25.5  \\ 
\hline \hline
\multirow{2}{*}{BP(317,\,309)} & \multirow{2}{*}{269\,fb} & \multirow{2}{*}{47996} & \multirow{2}{*}{45133} & \multirow{2}{*}{29750} &  195.1 & 131.2 & 92.0 \\
             &            &                    &                     &                     & (17.8,\,1.6) &  (17.7,\,2.4) & (18.3,\, 3.6)  \\
\multirow{2}{*}{BP(492,\,484)} & \multirow{2}{*}{32.7\,fb} & \multirow{2}{*}{5502} & \multirow{2}{*}{5131} & \multirow{2}{*}{3529} &  57.9 & 38.0 & 24.6 \\
             &            &                    &                     &                     & (5.3,\,0.48) &  (5.1,\,0.69) & (4.9,\,0.96)  \\
\multirow{2}{*}{BP(617,\,609)} & \multirow{2}{*}{9.56\,fb} & \multirow{2}{*}{1588} & \multirow{2}{*}{1471} & \multirow{2}{*}{1048} &  22.8 & 15.1 & 11.4 \\
             &            &                     &                     &                     & (2.1,\,0.19) &  (2.0,\,0.28) & (2.2,\,0.44)  \\
\hline             
\end{tabular}}
\caption{Number of signal ($\mathcal{S}$) and background events ($\mathcal{B}$) for the $\sqrt{s}=13$~TeV LHC with 
${\int \mathcal{L} \, dt =3}$~ab$^{-1}$ of integrated luminosity. We display the signal 
results for two benchmark points: $(m_{t^{(1)}},m_{X^{(1)}})/\gev=(492,484)$ 
and $(697,609)$, where $X^{(1)}\equiv \gamma^{(1)},Z^{(1)},W^{(1)},h^{(1)},\chi^{(1)}, H^{(1)\pm}$.
We show the signal sensitivity in brackets $(\mathcal{S}/\sqrt{\mathcal{B}},\mathcal{S}/\mathcal{B})$ in the last three columns. 
We consider all $t^{(1)}$, $T^{(1)}$, $B^{(1)}$ related modes. }
\label{tab:cutflow}
\end{table*}

We start our analysis requiring one isolated lepton  $p_{T\ell}>10$~GeV and $|\eta_\ell|<2.4$. Jets are 
defined via the anti-$k_T$ jet algorithm $R=0.4$, $p_{Tj}>20$~GeV and ${|\eta_j|<2.5}$ with the 
\textsc{FastJet} package~\cite{fastjet2}.  We require one $b$-jet with $b$-tagging  efficiency of 70\% that is 
associated to a mistag rate of 15\% for $c$-quarks and 1\% for light-quarks~\cite{btagging}. To tame the $\bar{t}t$ 
and $W\bar{b}b$ backgrounds, we explore the Jacobian peak structure for the signal, imposing
$m_{b\ell}<150$~GeV. The $W\bar{b}b$ background does not present this shape since it does not have 
a top-quark in the event and the $\bar{t}t$ typically produce a large tail, coming from events with the $b$ 
and $\ell$ combination from different top-quark decays. 

We further control the background with the transverse mass $m_T=\sqrt{2p_{T\ell}\slashed{E}_T(1-\cos\phi_{\ell,\slashed{E}_T})}$,
requiring $m_T>100$~GeV that explores the sharp drop above the $m_T\sim m_W$ for the  semi-leptonic $\bar{t}t$  
and $W\bar{b}b$ samples. \medskip

\begin{figure}[t!]
\includegraphics[width=1.15\columnwidth]{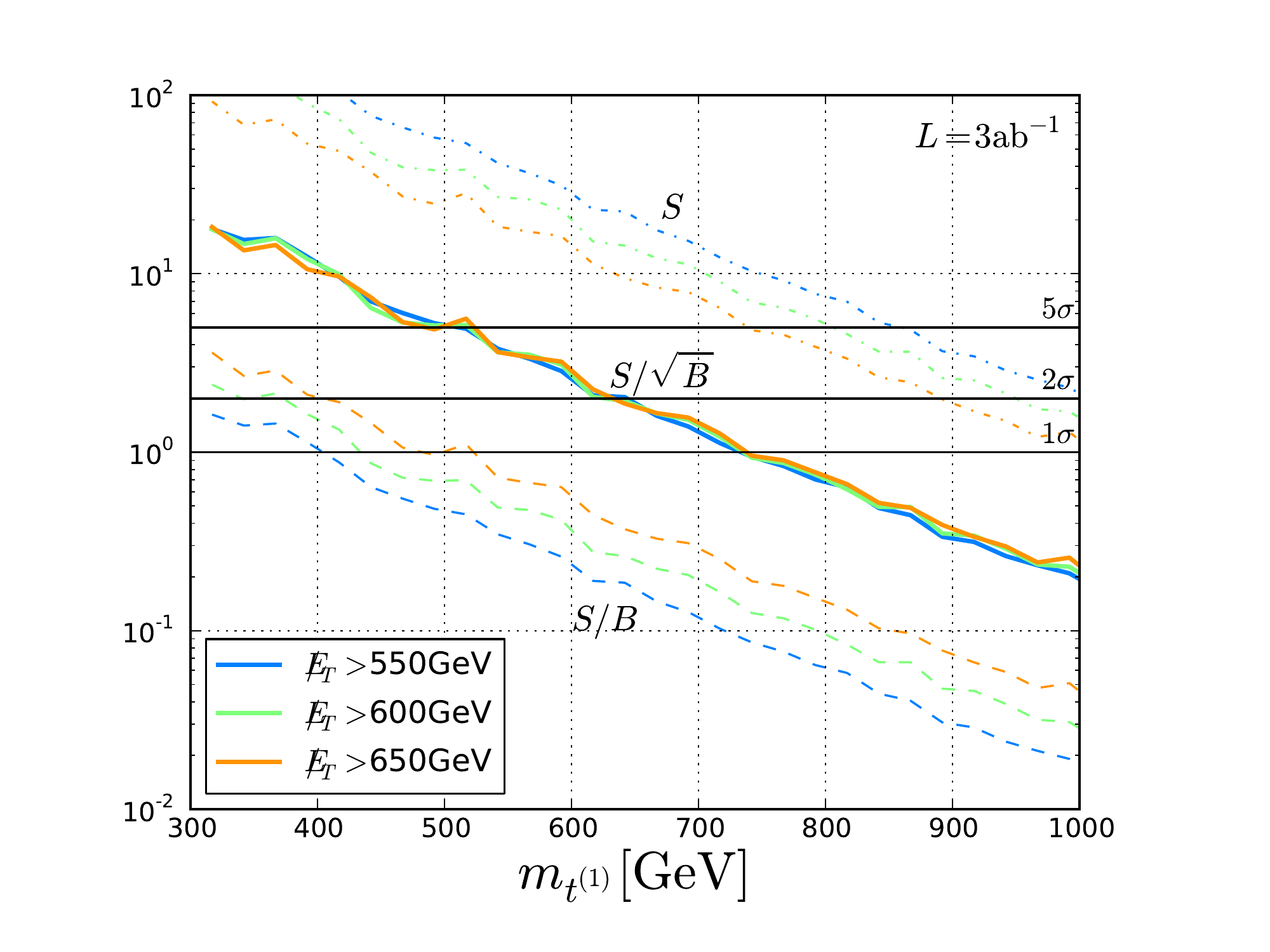}
\caption{Sensitivity lines $\mathcal{S}/\sqrt{\mathcal{B}}$ (solid)  and  $\mathcal{S}/\mathcal{B}$ (dashed) 
with $\int {\cal L} dt = 3$\,ab$^{-1}$
as functions of the top-partner mass $m_{t^{(1)}}$ in the 3rd generation extra dimensional model.
The mass splitting $m_{t^{(1)}} - m_{X^{(1)}}=8~\gev$ is assumed.
The results for different missing energy selections, $\slashed{E}_T/{\rm GeV} > 550$ (blue), 600 (green) and 650 (orange),
are shown.
 \label{fig:sensitivity}}
\end{figure}

In Fig.~\ref{fig:etmiss}, we present the resulting missing energy distributions for
signal and background for the 3rd generation extra dimensional model and the corresponding SUSY case. We set the top/bottom partner masses $m_{t^{(1)}}=m_{T^{(1)}}=m_{B^{(1)}}=317~\gev$.  
The signal distribution exhibits a lower suppression with $\slashed{E}_T$ compared to the backgrounds. 
We exploit this fact and define three signal regions with different requirements on the missing energy threshold,  
$\slashed{E}_T/\text{GeV}>550$, 600 and 650. The 
detailed signal and background cut-flow is displayed in table~\ref{tab:cutflow}. 
For scenario $(i)$, the main contribution comes from 
$B^{(1)} W_\mu^{(1)\pm} t$ sub-channel accounting for 28\% of the total rate, 
followed by 
$B^{(1)} H^{(1)\pm} t$ with 16\%, 
$T^{(1)} Z_\mu^{(1)} t$ with 14\%, 
$T^{(1)} h^{(1)} t$ with 8.8\%, 
$t^{(1)} \gamma_\mu^{(1)} t$  with 8.7\%, 
$T^{(1)} \chi^{(1)} t$ with 8.2\%,
$t^{(1)} h^{(1)} t$ with 8.0\%,  
$t^{(1)} \chi^{(1)} t$ with 7.9\% and  
$T^{(1)} \gamma_\mu^{(1)} t$ with 0.5\%. 
In summary, $B^{(1)}$, $T^{(1)}$, $t^{(1)}$ involved process, contributes 43\%, 32\%, 25\%, respectively.
This fractions are understood straightforwardly from the gauge couplings $g_1$, $g_2$ and top Yukawa coupling $\lambda_t$ and hypercharges. Here we ignore the processes whose amplitudes proportional to the bottom Yukawa coupling $\lambda_b$, as they present 
a sub-leading contribution.

In Fig.~\ref{fig:sensitivity}, we show $\mathcal{S}/\sqrt{\mathcal{B}}$ (solid lines), $\mathcal{S}/\mathcal{B}$ (dashed lines)
$\mathcal{S}$ (dotted lines)  as functions of the top-partner mass $m_{t^{(1)}}$ 
for scenario $(i)$, assuming the mass splitting ${m_{t^{(1)}} - m_{X^{(1)}}=8~\gev}$ and the 13~TeV LHC with 
${\int \mathcal{L} \, dt =3}$~ab$^{-1}$.  For completeness, we show the results for our different signal regions, ${\slashed{E}_T/{\rm GeV} > 550}$ 
(blue), 600 (green) and 650 (orange). They provide very similar sensitivities in $\mathcal{S}/\sqrt{\mathcal{B}}$. One can see that the top-partner 
mass can be probed up to ${m_{t^{(1)}}\sim 630~\gev}$ at 95\% CL in scenario $(i)$. 
A higher $\slashed{E}_T$ cut predicts smaller number of events but $\mathcal{S} \gtrsim 10$
can still be achieved with 3 ab$^{-1}$ around ${m_{t^{(1)}}\sim 630~\gev}$, while keeping ${\mathcal{S}/\mathcal{B} \gtrsim 0.3}$.

For the minimal top-partner simplified scenario $(ii)$, the sensitivity can be estimated 
by rescaling the cross section according to the contribution of the subprocesses quoted above, 
as we have checked that the missing energy distributions for all relevant processes are practically identical.
Since scenario $(ii)$ amounts to a signal that is purely $t^{(1)} h^{(1)} t$, its rate is tantamount  to only 8\,\% of the total rate in scenario $(i)$, as mentioned above. 
We find that the sensitivity reaches only just below 300~GeV in scenario $(ii)$. 
If we assume the signal strength of 2, corresponding for instance to an inclusion of the $t^{(1)} \chi^{(1)} t$ process, 
we find the 95\% CL expected limit on the top-partner mass ${m_{t^{(1)}} \gtrsim 400~\gev}$,
which should be directly compared with the sensitivity to the stop mass, ${m_{\tilde t} \gtrsim 380}~\gev$, in the natural SUSY scenario where two different channels corresponding to the two degenerate higgsino-like neutralinos ($\tilde \chi_1^0, \tilde \chi_2^0$) contribute.

 If the DM is selected to the spin-1 KK photon $\gamma^{(1)}_\mu$ instead of the KK Higgs $h^{(1)}$ in the minimal scenario, the strength of interaction is replaced by the $U(1)_Y$ gauge coupling multiplied by the hypercharge ${Y_R=4/3}$. As $\gamma^{(1)}_\mu$ has a larger degree of freedom compared to $h^{(1)}$, 
the cross-section of the $pp \to t^{(1)} \gamma^{(1)}_\mu t$ process appears to be almost identical 
to that of $pp \to t^{(1)} h^{(1)} t$.
We therefore have very similar conclusion for the ($t^{(1)}$, $\gamma^{(1)}_\mu$) minimal simplified scenario.

\medskip

\begin{figure}[t!]
\centering
\includegraphics[width=1.15\columnwidth]{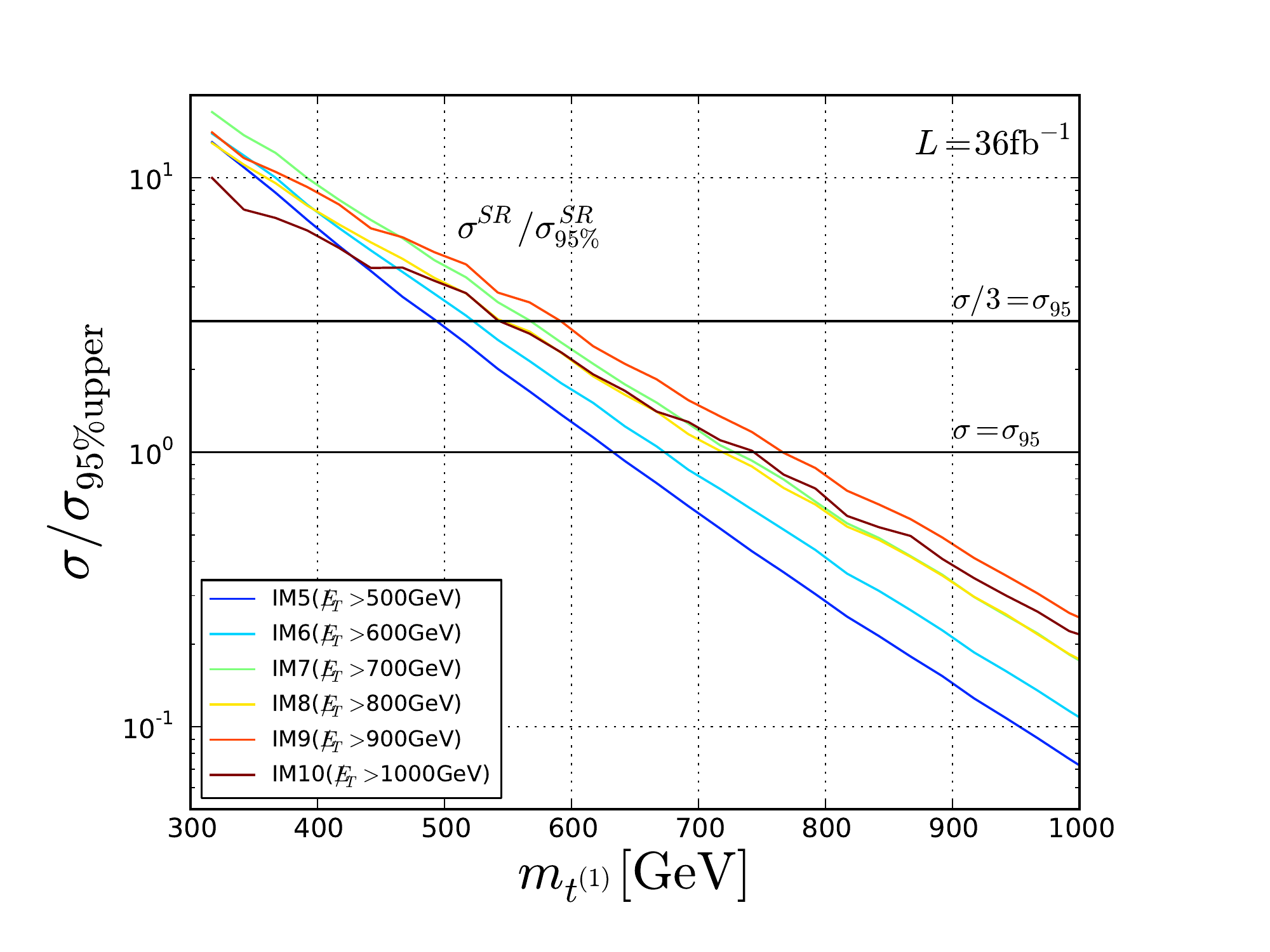}
\caption{Current bound on the mass of fermionic top-partner from the mono-jet study at 95\% C.L \cite{ATLAS:2017dnw}.
We display the results for different missing energy selections.
 \label{fig:monojet}}
\end{figure}

Finally, as in the SUSY scenario, mono-jet searches can provide important constraints for the degenerate spectrum.
We repeat similar analysis performed in Ref.~\cite{Choudhury:2016tff} with the KK tops ($t^{(1)},T^{(1)}$) and KK bottom ($B^{(1)}$) only. Since the work of Ref.~\cite{Choudhury:2016tff} includes KK gluon and all three generations of KK quarks in their mono-jet study,  their limit does not apply directly to our case, 
where only $t^{(1)}$, $T^{(1)}$, $B^{(1)}$ are considered.
We show current bound on the KK top mass in Fig.~\ref{fig:monojet}, using the data corresponding to an integrated luminosity of 36.1 fb$^{-1}$ at the 13 TeV LHC. We used the model-independent 95\% C.L. upper limits on signal cross section in the final state with an energetic jet and large missing transverse momentum reported by ATLAS~\cite{ATLAS:2017dnw}. Our analysis indicates that the current mono-jet study excludes the KK top mass up to $\sim 750$~GeV, which corresponds to $\sim 1.6$ TeV for a higher luminosity of 3 ab$^{-1}$ via a rough rescaling based on \cite{collider:reach}.
This is more powerful than the sensitivity of the mono-top channel, which implies that the mono-top channel is not a discovery channel and we should expect excesses both in the mono-jet and mono-top channels if we have a light top partner in the spectrum.
Significant improvements 
in the mono-top sensitivity can be obtained by also exploiting the hadronic mono-top final state~\cite{Goncalves:2016nil}.
%

\begin{figure}[t!]
\centering
\includegraphics[width=1.15\columnwidth]{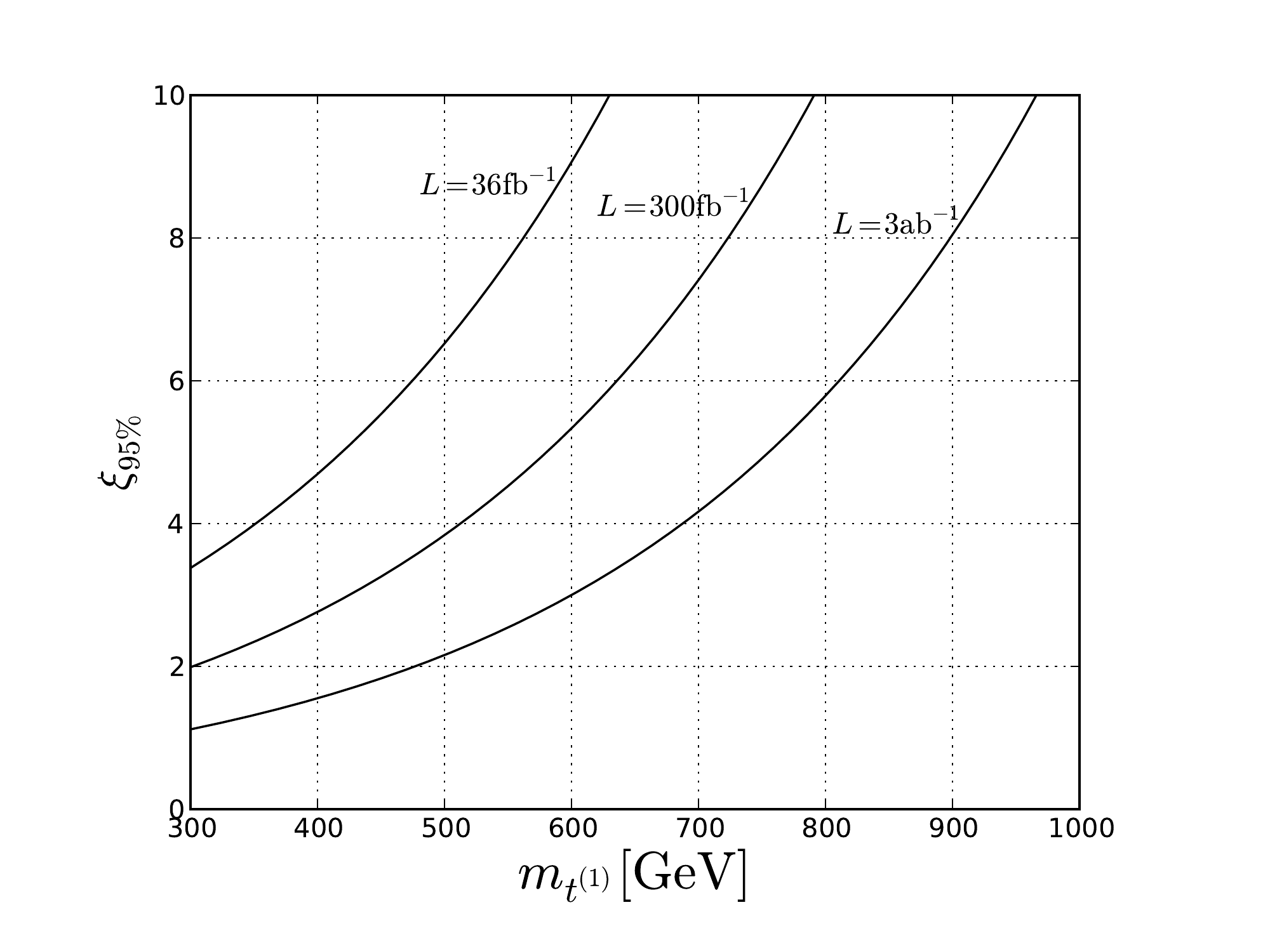}
\caption{The current and projected sensitivities on $\xi$ (introduced in Eq.~\eqref{eq:xi}).
The integrated luminosity of 36, 300 and 3000 fb$^{-1}$ are considered.
 \label{fig:xi}}
\end{figure}

It has been pointed out that the mono-top channel has a complementary role
to the mono-jet channel \cite{Goncalves:2016tft}.
What is observed in the mono-jet channel is the QCD initial state radiation
and the process carries too little information on the details of the top-partner and DM sectors.
Conversely, the existence of the top-quark in the mono-top channel is
a clear indication that the process is related to the 3rd generation.
The helicity of the top-quark can also be measured by looking at the angular distribution
between the charged lepton and the $b$-quark in the final state \cite{Goncalves:2016tft}, 
which provides important information on the chirality structure of the top-partner. 
Moreover, unlike the mono-jet channel, the production rate depends not only on the QCD coupling
but also on the couplings of new interactions involving the top-quark and the top-partner.
For example, in the simplified scenario $(ii)$, one can introduce the scaling factor $\xi$ as, 
\begin{equation}
{\cal L} \ni \xi\frac{ \lambda_t }{\sqrt{2}} \bar T_L t_R^{(1)}  h^{(1)} \,.
\label{eq:xi}
\end{equation}
With this parametrization, the signal strength scales as $\xi^2$ and one can set the limit on $\xi$
using the mono-top channel.
Using the previous analysis, we have estimated the current and projected sensitivities on $\xi$
for $\int {\cal L} dt = 36$, 300 and 3000 fb$^{-1}$ presented in Fig.~\ref{fig:xi}. 
One can see that for $e.g.$~$m_{t^{(1)}} \simeq 800$ GeV the high luminosity LHC can
prove $\xi$ up to around $6$.
The $\xi$ can also be effectively increased  
by introducing additional particles that couple to the top-quark and the top-partner (bottom-partner) 
in the same way as e.g.~$\gamma_\mu^{(1)}$ and $h^{(1)}$ ($W_\mu^{(1)}$ and $H^{(1)\pm}$).
If those new particles are only electroweakly charged, the enhancement of $\xi$ will be independent of the mono-jet channel. 
On the other hand, if they are colored, such as the KK gluon,
though the mono-top channel is significantly enhanced (due to {\it e.g.}~$pp \to t  t^{(1)}  g^{(1)}$ for the KK gluon case), the rate of mono-jet channel also increases due to the pair production
of those colored particles. 
Even though the sensitivity of the mono-top channel is in general weaker than that of the mono-jet channel,
it is important to look for the mono-top channel,
since this process provides important information on 
the top-partner and DM sectors in the fermionic top-partner models.

\section{Conclusion}
\label{sec:conclusion}

The prospects of observing the mono-top signatures at the LHC arising from fermionic top-partner models
have been studied.
Such a signature was previously studied in the $pp\rightarrow \tilde{t} t \tilde{h}^0$ process
in the context of Natural Supersymmetry, where the stop and higgsino present a very small mass gap, 
letting the $\tilde{t}$ decay products soft and undetectable~\cite{Goncalves:2016tft, Goncalves:2016nil}. 
Interestingly, a similar setup arises in the UED framework, where the compressed mass spectra are naturally expected.
In extra dimensional models many different channels contribute to the same mono-top final state, resulting in a large gain in 
the signal rate and the sensitivity. We showed that the mono-top channel can explore the top-partner 
masses up to 630~GeV (or 300~GeV in the simplified scenario) at the high luminosity LHC. 
Possible improvements in this bound can be obtained by taking also
the hadronic mono-top channel into account.
We have compared the mono-jet and mono-top channels and found that
the sensitivity of mono-jet channel is in general superior to the mono-top channel.
We have also argued that despite of weaker sensitivity of the mono-top channel,
it is important to observe and investigate this process
since it allows us to access the information on the fermionic top-partners
and the new particles that couple to the top-quark and the top-partner.
Since this channel has not been investigated experimentally  
in the contexts of supersymmetry and extra dimensional models,
we hope that more detailed studies will be performed by
the experimental collaborations at the LHC.

\section*{Acknowledgements}
KS thanks the TU Munich for hospitality during the final stages of this work and has been partially supported by the DFG cluster of excellence EXC 153 ``Origin and Structure of the Universe'', by the Collaborative Research Center SFB1258.
DG was funded by U.S. National Science Foundation under the grant PHY-1519175.
KK is supported in part by US Department of Energy under DE-SC0017965 and in part by the University of Kansas General Research Fund allocation \#2151081.
The work of KS is partially supported by the National Science Centre, Poland, under research grants
DEC-2014/15/B/ST2/02157, DEC-2015/18/M/ST2/00054, 2015/19/D/ST2/03136.
MT is supported in part by the JSPS Grant-in-Aid for
Scientific Research Numbers JP16H03991, JP16H02176, 17H05399,
and by World Premier International Research 
Center Initiative (WPI Initiative), MEXT, Japan. 

\bibliographystyle{apsrev}
\bibliography{draft}

\begin{thebibliography}{64}
\expandafter\ifx\csname natexlab\endcsname\relax\def\natexlab#1{#1}\fi
\expandafter\ifx\csname bibnamefont\endcsname\relax
  \def\bibnamefont#1{#1}\fi
\expandafter\ifx\csname bibfnamefont\endcsname\relax
  \def\bibfnamefont#1{#1}\fi
\expandafter\ifx\csname citenamefont\endcsname\relax
  \def\citenamefont#1{#1}\fi
\expandafter\ifx\csname url\endcsname\relax
  \def\url#1{\texttt{#1}}\fi
\expandafter\ifx\csname urlprefix\endcsname\relax\def\urlprefix{URL }\fi
\providecommand{\bibinfo}[2]{#2}
\providecommand{\eprint}[2][]{\url{#2}}

\bibitem[{\citenamefont{Goncalves et~al.}(2016)\citenamefont{Goncalves,
  Sakurai, and Takeuchi}}]{Goncalves:2016tft}
\bibinfo{author}{\bibfnamefont{D.}~\bibnamefont{Goncalves}},
  \bibinfo{author}{\bibfnamefont{K.}~\bibnamefont{Sakurai}}, \bibnamefont{and}
  \bibinfo{author}{\bibfnamefont{M.}~\bibnamefont{Takeuchi}},
  \bibinfo{journal}{Phys. Rev.} \textbf{\bibinfo{volume}{D94}},
  \bibinfo{pages}{075009} (\bibinfo{year}{2016}), \eprint{1604.03938}.

\bibitem[{\citenamefont{Goncalves et~al.}(2017)\citenamefont{Goncalves,
  Sakurai, and Takeuchi}}]{Goncalves:2016nil}
\bibinfo{author}{\bibfnamefont{D.}~\bibnamefont{Goncalves}},
  \bibinfo{author}{\bibfnamefont{K.}~\bibnamefont{Sakurai}}, \bibnamefont{and}
  \bibinfo{author}{\bibfnamefont{M.}~\bibnamefont{Takeuchi}},
  \bibinfo{journal}{Phys. Rev.} \textbf{\bibinfo{volume}{D95}},
  \bibinfo{pages}{015030} (\bibinfo{year}{2017}), \eprint{1610.06179}.

\bibitem[{\citenamefont{Fuks et~al.}(2015)\citenamefont{Fuks, Richardson, and
  Wilcock}}]{Fuks:2014lva}
\bibinfo{author}{\bibfnamefont{B.}~\bibnamefont{Fuks}},
  \bibinfo{author}{\bibfnamefont{P.}~\bibnamefont{Richardson}},
  \bibnamefont{and} \bibinfo{author}{\bibfnamefont{A.}~\bibnamefont{Wilcock}},
  \bibinfo{journal}{Eur. Phys. J.} \textbf{\bibinfo{volume}{C75}},
  \bibinfo{pages}{308} (\bibinfo{year}{2015}), \eprint{1408.3634}.

\bibitem[{\citenamefont{Aad et~al.}(2015)}]{Aad:2014wza}
\bibinfo{author}{\bibfnamefont{G.}~\bibnamefont{Aad}} \bibnamefont{et~al.}
  (\bibinfo{collaboration}{ATLAS}), \bibinfo{journal}{Eur. Phys. J.}
  \textbf{\bibinfo{volume}{C75}}, \bibinfo{pages}{79} (\bibinfo{year}{2015}),
  \eprint{1410.5404}.

\bibitem[{\citenamefont{Khachatryan et~al.}(2015)}]{Khachatryan:2014uma}
\bibinfo{author}{\bibfnamefont{V.}~\bibnamefont{Khachatryan}}
  \bibnamefont{et~al.} (\bibinfo{collaboration}{CMS}), \bibinfo{journal}{Phys.
  Rev. Lett.} \textbf{\bibinfo{volume}{114}}, \bibinfo{pages}{101801}
  (\bibinfo{year}{2015}), \eprint{1410.1149}.

\bibitem[{\citenamefont{Davoudiasl et~al.}(2011)\citenamefont{Davoudiasl,
  Morrissey, Sigurdson, and Tulin}}]{Davoudiasl:2011fj}
\bibinfo{author}{\bibfnamefont{H.}~\bibnamefont{Davoudiasl}},
  \bibinfo{author}{\bibfnamefont{D.~E.} \bibnamefont{Morrissey}},
  \bibinfo{author}{\bibfnamefont{K.}~\bibnamefont{Sigurdson}},
  \bibnamefont{and} \bibinfo{author}{\bibfnamefont{S.}~\bibnamefont{Tulin}},
  \bibinfo{journal}{Phys. Rev.} \textbf{\bibinfo{volume}{D84}},
  \bibinfo{pages}{096008} (\bibinfo{year}{2011}), \eprint{1106.4320}.

\bibitem[{\citenamefont{Kamenik and Zupan}(2011)}]{Kamenik:2011nb}
\bibinfo{author}{\bibfnamefont{J.~F.} \bibnamefont{Kamenik}} \bibnamefont{and}
  \bibinfo{author}{\bibfnamefont{J.}~\bibnamefont{Zupan}},
  \bibinfo{journal}{Phys. Rev.} \textbf{\bibinfo{volume}{D84}},
  \bibinfo{pages}{111502} (\bibinfo{year}{2011}), \eprint{1107.0623}.

\bibitem[{\citenamefont{Andrea et~al.}(2011)\citenamefont{Andrea, Fuks, and
  Maltoni}}]{Andrea:2011ws}
\bibinfo{author}{\bibfnamefont{J.}~\bibnamefont{Andrea}},
  \bibinfo{author}{\bibfnamefont{B.}~\bibnamefont{Fuks}}, \bibnamefont{and}
  \bibinfo{author}{\bibfnamefont{F.}~\bibnamefont{Maltoni}},
  \bibinfo{journal}{Phys. Rev.} \textbf{\bibinfo{volume}{D84}},
  \bibinfo{pages}{074025} (\bibinfo{year}{2011}), \eprint{1106.6199}.

\bibitem[{\citenamefont{Alvarez et~al.}(2014)\citenamefont{Alvarez, Leskow,
  Drobnak, and Kamenik}}]{Alvarez:2013jqa}
\bibinfo{author}{\bibfnamefont{E.}~\bibnamefont{Alvarez}},
  \bibinfo{author}{\bibfnamefont{E.~C.} \bibnamefont{Leskow}},
  \bibinfo{author}{\bibfnamefont{J.}~\bibnamefont{Drobnak}}, \bibnamefont{and}
  \bibinfo{author}{\bibfnamefont{J.~F.} \bibnamefont{Kamenik}},
  \bibinfo{journal}{Phys. Rev.} \textbf{\bibinfo{volume}{D89}},
  \bibinfo{pages}{014016} (\bibinfo{year}{2014}), \eprint{1310.7600}.

\bibitem[{\citenamefont{Agram et~al.}(2014)\citenamefont{Agram, Andrea,
  Buttignol, Conte, and Fuks}}]{Agram:2013wda}
\bibinfo{author}{\bibfnamefont{J.-L.} \bibnamefont{Agram}},
  \bibinfo{author}{\bibfnamefont{J.}~\bibnamefont{Andrea}},
  \bibinfo{author}{\bibfnamefont{M.}~\bibnamefont{Buttignol}},
  \bibinfo{author}{\bibfnamefont{E.}~\bibnamefont{Conte}}, \bibnamefont{and}
  \bibinfo{author}{\bibfnamefont{B.}~\bibnamefont{Fuks}},
  \bibinfo{journal}{Phys. Rev.} \textbf{\bibinfo{volume}{D89}},
  \bibinfo{pages}{014028} (\bibinfo{year}{2014}), \eprint{1311.6478}.

\bibitem[{\citenamefont{Boucheneb et~al.}(2015)\citenamefont{Boucheneb,
  Cacciapaglia, Deandrea, and Fuks}}]{Boucheneb:2014wza}
\bibinfo{author}{\bibfnamefont{I.}~\bibnamefont{Boucheneb}},
  \bibinfo{author}{\bibfnamefont{G.}~\bibnamefont{Cacciapaglia}},
  \bibinfo{author}{\bibfnamefont{A.}~\bibnamefont{Deandrea}}, \bibnamefont{and}
  \bibinfo{author}{\bibfnamefont{B.}~\bibnamefont{Fuks}},
  \bibinfo{journal}{JHEP} \textbf{\bibinfo{volume}{01}}, \bibinfo{pages}{017}
  (\bibinfo{year}{2015}), \eprint{1407.7529}.

\bibitem[{\citenamefont{Aaboud et~al.}(2016)}]{Aaboud:2016tnv}
\bibinfo{author}{\bibfnamefont{M.}~\bibnamefont{Aaboud}} \bibnamefont{et~al.}
  (\bibinfo{collaboration}{ATLAS}) (\bibinfo{year}{2016}), \eprint{1604.07773}.

\bibitem[{\citenamefont{Khachatryan et~al.}(2016)}]{Khachatryan:2016pxa}
\bibinfo{author}{\bibfnamefont{V.}~\bibnamefont{Khachatryan}}
  \bibnamefont{et~al.} (\bibinfo{collaboration}{CMS}),
  \bibinfo{journal}{Submitted to: Phys. Lett. B}  (\bibinfo{year}{2016}),
  \eprint{1605.08993}.

\bibitem[{\citenamefont{Kitano and Nomura}(2006)}]{Kitano:2006gv}
\bibinfo{author}{\bibfnamefont{R.}~\bibnamefont{Kitano}} \bibnamefont{and}
  \bibinfo{author}{\bibfnamefont{Y.}~\bibnamefont{Nomura}},
  \bibinfo{journal}{Phys. Rev.} \textbf{\bibinfo{volume}{D73}},
  \bibinfo{pages}{095004} (\bibinfo{year}{2006}), \eprint{hep-ph/0602096}.

\bibitem[{\citenamefont{Papucci et~al.}(2012)\citenamefont{Papucci, Ruderman,
  and Weiler}}]{Papucci:2011wy}
\bibinfo{author}{\bibfnamefont{M.}~\bibnamefont{Papucci}},
  \bibinfo{author}{\bibfnamefont{J.~T.} \bibnamefont{Ruderman}},
  \bibnamefont{and} \bibinfo{author}{\bibfnamefont{A.}~\bibnamefont{Weiler}},
  \bibinfo{journal}{JHEP} \textbf{\bibinfo{volume}{09}}, \bibinfo{pages}{035}
  (\bibinfo{year}{2012}), \eprint{1110.6926}.

\bibitem[{\citenamefont{Hall et~al.}(2012)\citenamefont{Hall, Pinner, and
  Ruderman}}]{Hall:2011aa}
\bibinfo{author}{\bibfnamefont{L.~J.} \bibnamefont{Hall}},
  \bibinfo{author}{\bibfnamefont{D.}~\bibnamefont{Pinner}}, \bibnamefont{and}
  \bibinfo{author}{\bibfnamefont{J.~T.} \bibnamefont{Ruderman}},
  \bibinfo{journal}{JHEP} \textbf{\bibinfo{volume}{04}}, \bibinfo{pages}{131}
  (\bibinfo{year}{2012}), \eprint{1112.2703}.

\bibitem[{\citenamefont{Desai and Mukhopadhyaya}(2012)}]{Desai:2011th}
\bibinfo{author}{\bibfnamefont{N.}~\bibnamefont{Desai}} \bibnamefont{and}
  \bibinfo{author}{\bibfnamefont{B.}~\bibnamefont{Mukhopadhyaya}},
  \bibinfo{journal}{JHEP} \textbf{\bibinfo{volume}{05}}, \bibinfo{pages}{057}
  (\bibinfo{year}{2012}), \eprint{1111.2830}.

\bibitem[{\citenamefont{Ishiwata et~al.}(2012)\citenamefont{Ishiwata, Nagata,
  and Yokozaki}}]{Ishiwata:2011ab}
\bibinfo{author}{\bibfnamefont{K.}~\bibnamefont{Ishiwata}},
  \bibinfo{author}{\bibfnamefont{N.}~\bibnamefont{Nagata}}, \bibnamefont{and}
  \bibinfo{author}{\bibfnamefont{N.}~\bibnamefont{Yokozaki}},
  \bibinfo{journal}{Phys. Lett.} \textbf{\bibinfo{volume}{B710}},
  \bibinfo{pages}{145} (\bibinfo{year}{2012}), \eprint{1112.1944}.

\bibitem[{\citenamefont{Sakurai and Takayama}(2011)}]{Sakurai:2011pt}
\bibinfo{author}{\bibfnamefont{K.}~\bibnamefont{Sakurai}} \bibnamefont{and}
  \bibinfo{author}{\bibfnamefont{K.}~\bibnamefont{Takayama}},
  \bibinfo{journal}{JHEP} \textbf{\bibinfo{volume}{12}}, \bibinfo{pages}{063}
  (\bibinfo{year}{2011}), \eprint{1106.3794}.

\bibitem[{\citenamefont{Kim et~al.}(2009)\citenamefont{Kim, Maekawa, Nagao,
  Nojiri, and Sakurai}}]{Kim:2009nq}
\bibinfo{author}{\bibfnamefont{S.-G.} \bibnamefont{Kim}},
  \bibinfo{author}{\bibfnamefont{N.}~\bibnamefont{Maekawa}},
  \bibinfo{author}{\bibfnamefont{K.~I.} \bibnamefont{Nagao}},
  \bibinfo{author}{\bibfnamefont{M.~M.} \bibnamefont{Nojiri}},
  \bibnamefont{and} \bibinfo{author}{\bibfnamefont{K.}~\bibnamefont{Sakurai}},
  \bibinfo{journal}{JHEP} \textbf{\bibinfo{volume}{10}}, \bibinfo{pages}{005}
  (\bibinfo{year}{2009}), \eprint{0907.4234}.

\bibitem[{\citenamefont{Wymant}(2012)}]{Wymant:2012zp}
\bibinfo{author}{\bibfnamefont{C.}~\bibnamefont{Wymant}},
  \bibinfo{journal}{Phys. Rev.} \textbf{\bibinfo{volume}{D86}},
  \bibinfo{pages}{115023} (\bibinfo{year}{2012}), \eprint{1208.1737}.

\bibitem[{\citenamefont{Baer et~al.}(2012{\natexlab{a}})\citenamefont{Baer,
  Barger, Huang, Mustafayev, and Tata}}]{Baer:2012up}
\bibinfo{author}{\bibfnamefont{H.}~\bibnamefont{Baer}},
  \bibinfo{author}{\bibfnamefont{V.}~\bibnamefont{Barger}},
  \bibinfo{author}{\bibfnamefont{P.}~\bibnamefont{Huang}},
  \bibinfo{author}{\bibfnamefont{A.}~\bibnamefont{Mustafayev}},
  \bibnamefont{and} \bibinfo{author}{\bibfnamefont{X.}~\bibnamefont{Tata}},
  \bibinfo{journal}{Phys. Rev. Lett.} \textbf{\bibinfo{volume}{109}},
  \bibinfo{pages}{161802} (\bibinfo{year}{2012}{\natexlab{a}}),
  \eprint{1207.3343}.

\bibitem[{\citenamefont{Randall and Reece}(2013)}]{Randall:2012dm}
\bibinfo{author}{\bibfnamefont{L.}~\bibnamefont{Randall}} \bibnamefont{and}
  \bibinfo{author}{\bibfnamefont{M.}~\bibnamefont{Reece}},
  \bibinfo{journal}{JHEP} \textbf{\bibinfo{volume}{08}}, \bibinfo{pages}{088}
  (\bibinfo{year}{2013}), \eprint{1206.6540}.

\bibitem[{\citenamefont{Cao et~al.}(2012)\citenamefont{Cao, Han, Wu, Yang, and
  Zhang}}]{Cao:2012rz}
\bibinfo{author}{\bibfnamefont{J.}~\bibnamefont{Cao}},
  \bibinfo{author}{\bibfnamefont{C.}~\bibnamefont{Han}},
  \bibinfo{author}{\bibfnamefont{L.}~\bibnamefont{Wu}},
  \bibinfo{author}{\bibfnamefont{J.~M.} \bibnamefont{Yang}}, \bibnamefont{and}
  \bibinfo{author}{\bibfnamefont{Y.}~\bibnamefont{Zhang}},
  \bibinfo{journal}{JHEP} \textbf{\bibinfo{volume}{11}}, \bibinfo{pages}{039}
  (\bibinfo{year}{2012}), \eprint{1206.3865}.

\bibitem[{\citenamefont{Asano and Higaki}(2012)}]{Asano:2012sv}
\bibinfo{author}{\bibfnamefont{M.}~\bibnamefont{Asano}} \bibnamefont{and}
  \bibinfo{author}{\bibfnamefont{T.}~\bibnamefont{Higaki}},
  \bibinfo{journal}{Phys. Rev.} \textbf{\bibinfo{volume}{D86}},
  \bibinfo{pages}{035020} (\bibinfo{year}{2012}), \eprint{1204.0508}.

\bibitem[{\citenamefont{Baer et~al.}(2012{\natexlab{b}})\citenamefont{Baer,
  Barger, Huang, and Tata}}]{Baer:2012uy}
\bibinfo{author}{\bibfnamefont{H.}~\bibnamefont{Baer}},
  \bibinfo{author}{\bibfnamefont{V.}~\bibnamefont{Barger}},
  \bibinfo{author}{\bibfnamefont{P.}~\bibnamefont{Huang}}, \bibnamefont{and}
  \bibinfo{author}{\bibfnamefont{X.}~\bibnamefont{Tata}},
  \bibinfo{journal}{JHEP} \textbf{\bibinfo{volume}{05}}, \bibinfo{pages}{109}
  (\bibinfo{year}{2012}{\natexlab{b}}), \eprint{1203.5539}.

\bibitem[{\citenamefont{Evans et~al.}(2014)\citenamefont{Evans, Kats, Shih, and
  Strassler}}]{Evans:2013jna}
\bibinfo{author}{\bibfnamefont{J.~A.} \bibnamefont{Evans}},
  \bibinfo{author}{\bibfnamefont{Y.}~\bibnamefont{Kats}},
  \bibinfo{author}{\bibfnamefont{D.}~\bibnamefont{Shih}}, \bibnamefont{and}
  \bibinfo{author}{\bibfnamefont{M.~J.} \bibnamefont{Strassler}},
  \bibinfo{journal}{JHEP} \textbf{\bibinfo{volume}{07}}, \bibinfo{pages}{101}
  (\bibinfo{year}{2014}), \eprint{1310.5758}.

\bibitem[{\citenamefont{Hardy}(2013)}]{Hardy:2013ywa}
\bibinfo{author}{\bibfnamefont{E.}~\bibnamefont{Hardy}},
  \bibinfo{journal}{JHEP} \textbf{\bibinfo{volume}{10}}, \bibinfo{pages}{133}
  (\bibinfo{year}{2013}), \eprint{1306.1534}.

\bibitem[{\citenamefont{Kribs et~al.}(2013)\citenamefont{Kribs, Martin, and
  Menon}}]{Kribs:2013lua}
\bibinfo{author}{\bibfnamefont{G.~D.} \bibnamefont{Kribs}},
  \bibinfo{author}{\bibfnamefont{A.}~\bibnamefont{Martin}}, \bibnamefont{and}
  \bibinfo{author}{\bibfnamefont{A.}~\bibnamefont{Menon}},
  \bibinfo{journal}{Phys. Rev.} \textbf{\bibinfo{volume}{D88}},
  \bibinfo{pages}{035025} (\bibinfo{year}{2013}), \eprint{1305.1313}.

\bibitem[{\citenamefont{Bhattacherjee et~al.}(2013)\citenamefont{Bhattacherjee,
  Evans, Ibe, Matsumoto, and Yanagida}}]{Bhattacherjee:2013gr}
\bibinfo{author}{\bibfnamefont{B.}~\bibnamefont{Bhattacherjee}},
  \bibinfo{author}{\bibfnamefont{J.~L.} \bibnamefont{Evans}},
  \bibinfo{author}{\bibfnamefont{M.}~\bibnamefont{Ibe}},
  \bibinfo{author}{\bibfnamefont{S.}~\bibnamefont{Matsumoto}},
  \bibnamefont{and} \bibinfo{author}{\bibfnamefont{T.~T.}
  \bibnamefont{Yanagida}}, \bibinfo{journal}{Phys. Rev.}
  \textbf{\bibinfo{volume}{D87}}, \bibinfo{pages}{115002}
  (\bibinfo{year}{2013}), \eprint{1301.2336}.

\bibitem[{\citenamefont{Rolbiecki and Sakurai}(2013)}]{Rolbiecki:2013fia}
\bibinfo{author}{\bibfnamefont{K.}~\bibnamefont{Rolbiecki}} \bibnamefont{and}
  \bibinfo{author}{\bibfnamefont{K.}~\bibnamefont{Sakurai}},
  \bibinfo{journal}{JHEP} \textbf{\bibinfo{volume}{09}}, \bibinfo{pages}{004}
  (\bibinfo{year}{2013}), \eprint{1303.5696}.

\bibitem[{\citenamefont{Appelquist et~al.}(2001)\citenamefont{Appelquist,
  Cheng, and Dobrescu}}]{Appelquist:2000nn}
\bibinfo{author}{\bibfnamefont{T.}~\bibnamefont{Appelquist}},
  \bibinfo{author}{\bibfnamefont{H.-C.} \bibnamefont{Cheng}}, \bibnamefont{and}
  \bibinfo{author}{\bibfnamefont{B.~A.} \bibnamefont{Dobrescu}},
  \bibinfo{journal}{Phys. Rev.} \textbf{\bibinfo{volume}{D64}},
  \bibinfo{pages}{035002} (\bibinfo{year}{2001}), \eprint{hep-ph/0012100}.

\bibitem[{\citenamefont{Georgi et~al.}(2001)\citenamefont{Georgi, Grant, and
  Hailu}}]{Georgi:2000wb}
\bibinfo{author}{\bibfnamefont{H.}~\bibnamefont{Georgi}},
  \bibinfo{author}{\bibfnamefont{A.~K.} \bibnamefont{Grant}}, \bibnamefont{and}
  \bibinfo{author}{\bibfnamefont{G.}~\bibnamefont{Hailu}},
  \bibinfo{journal}{Phys. Rev.} \textbf{\bibinfo{volume}{D63}},
  \bibinfo{pages}{064027} (\bibinfo{year}{2001}), \eprint{hep-ph/0007350}.

\bibitem[{\citenamefont{Cheng et~al.}(2002{\natexlab{a}})\citenamefont{Cheng,
  Matchev, and Schmaltz}}]{Cheng:2002ab}
\bibinfo{author}{\bibfnamefont{H.-C.} \bibnamefont{Cheng}},
  \bibinfo{author}{\bibfnamefont{K.~T.} \bibnamefont{Matchev}},
  \bibnamefont{and} \bibinfo{author}{\bibfnamefont{M.}~\bibnamefont{Schmaltz}},
  \bibinfo{journal}{Phys. Rev.} \textbf{\bibinfo{volume}{D66}},
  \bibinfo{pages}{056006} (\bibinfo{year}{2002}{\natexlab{a}}),
  \eprint{hep-ph/0205314}.

\bibitem[{\citenamefont{Cheng et~al.}(2002{\natexlab{b}})\citenamefont{Cheng,
  Matchev, and Schmaltz}}]{Cheng:2002iz}
\bibinfo{author}{\bibfnamefont{H.-C.} \bibnamefont{Cheng}},
  \bibinfo{author}{\bibfnamefont{K.~T.} \bibnamefont{Matchev}},
  \bibnamefont{and} \bibinfo{author}{\bibfnamefont{M.}~\bibnamefont{Schmaltz}},
  \bibinfo{journal}{Phys. Rev.} \textbf{\bibinfo{volume}{D66}},
  \bibinfo{pages}{036005} (\bibinfo{year}{2002}{\natexlab{b}}),
  \eprint{hep-ph/0204342}.

\bibitem[{\citenamefont{Flacke et~al.}(2013)\citenamefont{Flacke, Kong, and
  Park}}]{Flacke:2013pla}
\bibinfo{author}{\bibfnamefont{T.}~\bibnamefont{Flacke}},
  \bibinfo{author}{\bibfnamefont{K.}~\bibnamefont{Kong}}, \bibnamefont{and}
  \bibinfo{author}{\bibfnamefont{S.~C.} \bibnamefont{Park}},
  \bibinfo{journal}{JHEP} \textbf{\bibinfo{volume}{05}}, \bibinfo{pages}{111}
  (\bibinfo{year}{2013}), \eprint{1303.0872}.

\bibitem[{\citenamefont{Flacke et~al.}(2017)\citenamefont{Flacke, Kang, Kong,
  Mohlabeng, and Park}}]{Flacke:2017xsv}
\bibinfo{author}{\bibfnamefont{T.}~\bibnamefont{Flacke}},
  \bibinfo{author}{\bibfnamefont{D.~W.} \bibnamefont{Kang}},
  \bibinfo{author}{\bibfnamefont{K.}~\bibnamefont{Kong}},
  \bibinfo{author}{\bibfnamefont{G.}~\bibnamefont{Mohlabeng}},
  \bibnamefont{and} \bibinfo{author}{\bibfnamefont{S.~C.} \bibnamefont{Park}},
  \bibinfo{journal}{JHEP} \textbf{\bibinfo{volume}{04}}, \bibinfo{pages}{041}
  (\bibinfo{year}{2017}), \eprint{1702.02949}.

\bibitem[{\citenamefont{Choudhury and Ghosh}(2016)}]{Choudhury:2016tff}
\bibinfo{author}{\bibfnamefont{D.}~\bibnamefont{Choudhury}} \bibnamefont{and}
  \bibinfo{author}{\bibfnamefont{K.}~\bibnamefont{Ghosh}},
  \bibinfo{journal}{Phys. Lett.} \textbf{\bibinfo{volume}{B763}},
  \bibinfo{pages}{155} (\bibinfo{year}{2016}), \eprint{1606.04084}.

\bibitem[{\citenamefont{Deutschmann et~al.}(2017)\citenamefont{Deutschmann,
  Flacke, and Kim}}]{Deutschmann:2017bth}
\bibinfo{author}{\bibfnamefont{N.}~\bibnamefont{Deutschmann}},
  \bibinfo{author}{\bibfnamefont{T.}~\bibnamefont{Flacke}}, \bibnamefont{and}
  \bibinfo{author}{\bibfnamefont{J.~S.} \bibnamefont{Kim}},
  \bibinfo{journal}{Phys. Lett.} \textbf{\bibinfo{volume}{B771}},
  \bibinfo{pages}{515} (\bibinfo{year}{2017}), \eprint{1702.00410}.

\bibitem[{\citenamefont{Beuria et~al.}(2017)\citenamefont{Beuria, Datta,
  Debnath, and Matchev}}]{Beuria:2017jez}
\bibinfo{author}{\bibfnamefont{J.}~\bibnamefont{Beuria}},
  \bibinfo{author}{\bibfnamefont{A.}~\bibnamefont{Datta}},
  \bibinfo{author}{\bibfnamefont{D.}~\bibnamefont{Debnath}}, \bibnamefont{and}
  \bibinfo{author}{\bibfnamefont{K.~T.} \bibnamefont{Matchev}}
  (\bibinfo{year}{2017}), \eprint{1702.00413}.

\bibitem[{\citenamefont{Belanger et~al.}(2011)\citenamefont{Belanger, Kakizaki,
  and Pukhov}}]{Belanger:2010yx}
\bibinfo{author}{\bibfnamefont{G.}~\bibnamefont{Belanger}},
  \bibinfo{author}{\bibfnamefont{M.}~\bibnamefont{Kakizaki}}, \bibnamefont{and}
  \bibinfo{author}{\bibfnamefont{A.}~\bibnamefont{Pukhov}},
  \bibinfo{journal}{JCAP} \textbf{\bibinfo{volume}{1102}}, \bibinfo{pages}{009}
  (\bibinfo{year}{2011}), \eprint{1012.2577}.

\bibitem[{\citenamefont{Servant and Tait}(2002)}]{Servant:2002hb}
\bibinfo{author}{\bibfnamefont{G.}~\bibnamefont{Servant}} \bibnamefont{and}
  \bibinfo{author}{\bibfnamefont{T.~M.~P.} \bibnamefont{Tait}},
  \bibinfo{journal}{New J. Phys.} \textbf{\bibinfo{volume}{4}},
  \bibinfo{pages}{99} (\bibinfo{year}{2002}), \eprint{hep-ph/0209262}.

\bibitem[{\citenamefont{Kong and Matchev}(2006)}]{Kong:2005hn}
\bibinfo{author}{\bibfnamefont{K.}~\bibnamefont{Kong}} \bibnamefont{and}
  \bibinfo{author}{\bibfnamefont{K.~T.} \bibnamefont{Matchev}},
  \bibinfo{journal}{JHEP} \textbf{\bibinfo{volume}{01}}, \bibinfo{pages}{038}
  (\bibinfo{year}{2006}), \eprint{hep-ph/0509119}.

\bibitem[{\citenamefont{Cheng et~al.}(2002{\natexlab{c}})\citenamefont{Cheng,
  Feng, and Matchev}}]{Cheng:2002ej}
\bibinfo{author}{\bibfnamefont{H.-C.} \bibnamefont{Cheng}},
  \bibinfo{author}{\bibfnamefont{J.~L.} \bibnamefont{Feng}}, \bibnamefont{and}
  \bibinfo{author}{\bibfnamefont{K.~T.} \bibnamefont{Matchev}},
  \bibinfo{journal}{Phys. Rev. Lett.} \textbf{\bibinfo{volume}{89}},
  \bibinfo{pages}{211301} (\bibinfo{year}{2002}{\natexlab{c}}),
  \eprint{hep-ph/0207125}.

\bibitem[{\citenamefont{Arrenberg et~al.}(2008)\citenamefont{Arrenberg, Baudis,
  Kong, Matchev, and Yoo}}]{Arrenberg:2008wy}
\bibinfo{author}{\bibfnamefont{S.}~\bibnamefont{Arrenberg}},
  \bibinfo{author}{\bibfnamefont{L.}~\bibnamefont{Baudis}},
  \bibinfo{author}{\bibfnamefont{K.}~\bibnamefont{Kong}},
  \bibinfo{author}{\bibfnamefont{K.~T.} \bibnamefont{Matchev}},
  \bibnamefont{and} \bibinfo{author}{\bibfnamefont{J.}~\bibnamefont{Yoo}},
  \bibinfo{journal}{Phys. Rev.} \textbf{\bibinfo{volume}{D78}},
  \bibinfo{pages}{056002} (\bibinfo{year}{2008}), \eprint{0805.4210}.

\bibitem[{\citenamefont{Datta et~al.}(2005)\citenamefont{Datta, Kong, and
  Matchev}}]{Datta:2005zs}
\bibinfo{author}{\bibfnamefont{A.}~\bibnamefont{Datta}},
  \bibinfo{author}{\bibfnamefont{K.}~\bibnamefont{Kong}}, \bibnamefont{and}
  \bibinfo{author}{\bibfnamefont{K.~T.} \bibnamefont{Matchev}},
  \bibinfo{journal}{Phys. Rev.} \textbf{\bibinfo{volume}{D72}},
  \bibinfo{pages}{096006} (\bibinfo{year}{2005}), \bibinfo{note}{[Erratum:
  Phys. Rev.D72,119901(2005)]}, \eprint{hep-ph/0509246}.

\bibitem[{ATL(2017{\natexlab{a}})}]{ATLAS-CONF-2017-022}
\bibinfo{type}{Tech. Rep.} \bibinfo{number}{ATLAS-CONF-2017-022},
  \bibinfo{institution}{CERN}, \bibinfo{address}{Geneva}
  (\bibinfo{year}{2017}{\natexlab{a}}),
  \urlprefix\url{https://cds.cern.ch/record/2258145}.

\bibitem[{ATL(2017{\natexlab{b}})}]{ATLAS-CONF-2017-060}
\bibinfo{type}{Tech. Rep.} \bibinfo{number}{ATLAS-CONF-2017-060},
  \bibinfo{institution}{CERN}, \bibinfo{address}{Geneva}
  (\bibinfo{year}{2017}{\natexlab{b}}),
  \urlprefix\url{http://cds.cern.ch/record/2273876}.

\bibitem[{\citenamefont{Datta and Shaw}(2017)}]{Datta:2016flx}
\bibinfo{author}{\bibfnamefont{A.}~\bibnamefont{Datta}} \bibnamefont{and}
  \bibinfo{author}{\bibfnamefont{A.}~\bibnamefont{Shaw}}
  (\bibinfo{collaboration}{Indian Association for the Cultivation of Science}),
  \bibinfo{journal}{Phys. Rev.} \textbf{\bibinfo{volume}{D95}},
  \bibinfo{pages}{015033} (\bibinfo{year}{2017}), \eprint{1610.09924}.

\bibitem[{\citenamefont{Biswas et~al.}(2017)\citenamefont{Biswas, Patra, and
  Shaw}}]{Biswas:2017vhc}
\bibinfo{author}{\bibfnamefont{A.}~\bibnamefont{Biswas}},
  \bibinfo{author}{\bibfnamefont{S.~K.} \bibnamefont{Patra}}, \bibnamefont{and}
  \bibinfo{author}{\bibfnamefont{A.}~\bibnamefont{Shaw}}
  (\bibinfo{year}{2017}), \eprint{1708.08938}.

\bibitem[{\citenamefont{Flacke et~al.}(2009)\citenamefont{Flacke, Menon, and
  Phalen}}]{Flacke:2008ne}
\bibinfo{author}{\bibfnamefont{T.}~\bibnamefont{Flacke}},
  \bibinfo{author}{\bibfnamefont{A.}~\bibnamefont{Menon}}, \bibnamefont{and}
  \bibinfo{author}{\bibfnamefont{D.~J.} \bibnamefont{Phalen}},
  \bibinfo{journal}{Phys. Rev.} \textbf{\bibinfo{volume}{D79}},
  \bibinfo{pages}{056009} (\bibinfo{year}{2009}), \eprint{0811.1598}.

\bibitem[{\citenamefont{Ishigure et~al.}(2016)\citenamefont{Ishigure, Kakizaki,
  and Santa}}]{Ishigure:2016kxp}
\bibinfo{author}{\bibfnamefont{Y.}~\bibnamefont{Ishigure}},
  \bibinfo{author}{\bibfnamefont{M.}~\bibnamefont{Kakizaki}}, \bibnamefont{and}
  \bibinfo{author}{\bibfnamefont{A.}~\bibnamefont{Santa}}
  (\bibinfo{year}{2016}), \eprint{1611.06760}.

\bibitem[{\citenamefont{Huang et~al.}(2012)\citenamefont{Huang, Kong, and
  Park}}]{Huang:2012kz}
\bibinfo{author}{\bibfnamefont{G.-Y.} \bibnamefont{Huang}},
  \bibinfo{author}{\bibfnamefont{K.}~\bibnamefont{Kong}}, \bibnamefont{and}
  \bibinfo{author}{\bibfnamefont{S.~C.} \bibnamefont{Park}},
  \bibinfo{journal}{JHEP} \textbf{\bibinfo{volume}{06}}, \bibinfo{pages}{099}
  (\bibinfo{year}{2012}), \eprint{1204.0522}.

\bibitem[{\citenamefont{Mangano et~al.}(2003)\citenamefont{Mangano, Moretti,
  Piccinini, Pittau, and Polosa}}]{Mangano:2002ea}
\bibinfo{author}{\bibfnamefont{M.~L.} \bibnamefont{Mangano}},
  \bibinfo{author}{\bibfnamefont{M.}~\bibnamefont{Moretti}},
  \bibinfo{author}{\bibfnamefont{F.}~\bibnamefont{Piccinini}},
  \bibinfo{author}{\bibfnamefont{R.}~\bibnamefont{Pittau}}, \bibnamefont{and}
  \bibinfo{author}{\bibfnamefont{A.~D.} \bibnamefont{Polosa}},
  \bibinfo{journal}{JHEP} \textbf{\bibinfo{volume}{07}}, \bibinfo{pages}{001}
  (\bibinfo{year}{2003}), \eprint{hep-ph/0206293}.

\bibitem[{\citenamefont{Alwall et~al.}(2014)\citenamefont{Alwall, Frederix,
  Frixione, Hirschi, Maltoni, Mattelaer, Shao, Stelzer, Torrielli, and
  Zaro}}]{mg5}
\bibinfo{author}{\bibfnamefont{J.}~\bibnamefont{Alwall}},
  \bibinfo{author}{\bibfnamefont{R.}~\bibnamefont{Frederix}},
  \bibinfo{author}{\bibfnamefont{S.}~\bibnamefont{Frixione}},
  \bibinfo{author}{\bibfnamefont{V.}~\bibnamefont{Hirschi}},
  \bibinfo{author}{\bibfnamefont{F.}~\bibnamefont{Maltoni}},
  \bibinfo{author}{\bibfnamefont{O.}~\bibnamefont{Mattelaer}},
  \bibinfo{author}{\bibfnamefont{H.~S.} \bibnamefont{Shao}},
  \bibinfo{author}{\bibfnamefont{T.}~\bibnamefont{Stelzer}},
  \bibinfo{author}{\bibfnamefont{P.}~\bibnamefont{Torrielli}},
  \bibnamefont{and} \bibinfo{author}{\bibfnamefont{M.}~\bibnamefont{Zaro}},
  \bibinfo{journal}{JHEP} \textbf{\bibinfo{volume}{07}}, \bibinfo{pages}{079}
  (\bibinfo{year}{2014}), \eprint{1405.0301}.

\bibitem[{\citenamefont{Sjostrand et~al.}(2008)\citenamefont{Sjostrand, Mrenna,
  and Skands}}]{pythia}
\bibinfo{author}{\bibfnamefont{T.}~\bibnamefont{Sjostrand}},
  \bibinfo{author}{\bibfnamefont{S.}~\bibnamefont{Mrenna}}, \bibnamefont{and}
  \bibinfo{author}{\bibfnamefont{P.~Z.} \bibnamefont{Skands}},
  \bibinfo{journal}{Comput. Phys. Commun.} \textbf{\bibinfo{volume}{178}},
  \bibinfo{pages}{852} (\bibinfo{year}{2008}), \eprint{0710.3820}.

\bibitem[{\citenamefont{Ovyn et~al.}(2009)\citenamefont{Ovyn, Rouby, and
  Lemaitre}}]{delphes}
\bibinfo{author}{\bibfnamefont{S.}~\bibnamefont{Ovyn}},
  \bibinfo{author}{\bibfnamefont{X.}~\bibnamefont{Rouby}}, \bibnamefont{and}
  \bibinfo{author}{\bibfnamefont{V.}~\bibnamefont{Lemaitre}}
  (\bibinfo{year}{2009}), \eprint{0903.2225}.

\bibitem[{\citenamefont{Czakon et~al.}(2013)\citenamefont{Czakon, Fiedler, and
  Mitov}}]{tt_NNLO}
\bibinfo{author}{\bibfnamefont{M.}~\bibnamefont{Czakon}},
  \bibinfo{author}{\bibfnamefont{P.}~\bibnamefont{Fiedler}}, \bibnamefont{and}
  \bibinfo{author}{\bibfnamefont{A.}~\bibnamefont{Mitov}},
  \bibinfo{journal}{Phys. Rev. Lett.} \textbf{\bibinfo{volume}{110}},
  \bibinfo{pages}{252004} (\bibinfo{year}{2013}), \eprint{1303.6254}.

\bibitem[{\citenamefont{{LHC Top Working Group:}}(2016)}]{LHCTopWG}
\bibinfo{author}{\bibnamefont{{LHC Top Working Group:}}}
  (\bibinfo{year}{2016}),
  \eprint{https://twiki.cern.ch/twiki/bin/view/LHCPhysics/ SingleTopRefXsec}.

\bibitem[{\citenamefont{Campbell et~al.}(2013)\citenamefont{Campbell, Ellis,
  and Rontsch}}]{tz}
\bibinfo{author}{\bibfnamefont{J.}~\bibnamefont{Campbell}},
  \bibinfo{author}{\bibfnamefont{R.~K.} \bibnamefont{Ellis}}, \bibnamefont{and}
  \bibinfo{author}{\bibfnamefont{R.}~\bibnamefont{Rontsch}},
  \bibinfo{journal}{Phys. Rev.} \textbf{\bibinfo{volume}{D87}},
  \bibinfo{pages}{114006} (\bibinfo{year}{2013}), \eprint{1302.3856}.

\bibitem[{\citenamefont{Cacciari et~al.}(2012)\citenamefont{Cacciari, Salam,
  and Soyez}}]{fastjet2}
\bibinfo{author}{\bibfnamefont{M.}~\bibnamefont{Cacciari}},
  \bibinfo{author}{\bibfnamefont{G.~P.} \bibnamefont{Salam}}, \bibnamefont{and}
  \bibinfo{author}{\bibfnamefont{G.}~\bibnamefont{Soyez}},
  \bibinfo{journal}{Eur. Phys. J.} \textbf{\bibinfo{volume}{C72}},
  \bibinfo{pages}{1896} (\bibinfo{year}{2012}), \eprint{1111.6097}.

\bibitem[{\citenamefont{CMS\hspace*{0.1cm}Collaboration}(2013)}]{btagging}
\bibinfo{author}{\bibnamefont{CMS\hspace*{0.1cm}Collaboration}},
  \bibinfo{journal}{CMS-PAS-BTV-13-001}  (\bibinfo{year}{2013}).

\bibitem[{\citenamefont{ATLAS\hspace*{0.1cm}Collaboration}(2017)}]{ATLAS:2017dnw}
\bibinfo{author}{\bibnamefont{ATLAS\hspace*{0.1cm}Collaboration}}
  (\bibinfo{collaboration}{ATLAS}), \bibinfo{journal}{ATLAS-CONF-2017-060}
  (\bibinfo{year}{2017}).

\bibitem[{\citenamefont{Reach($\beta$)}(2014)}]{collider:reach}
\bibinfo{author}{\bibfnamefont{C.}~\bibnamefont{Reach($\beta$)}},
  \bibinfo{journal}{{http://collider-reach.web.cern.ch}}
  (\bibinfo{year}{2014}).

\end{thebibliography}

\end{document}